\documentclass[twocolumn,superscriptaddress,aps,reprint, amsmath,amssymb, prapplied,longbibliography]{revtex4-2}
\usepackage{xfrac}
\usepackage{graphicx}
\usepackage{geometry}
\usepackage[inkscapelatex=false]{svg}
\usepackage{multirow}
\usepackage{amsmath,amssymb,amsfonts}
\usepackage{amsthm}
\usepackage{mathrsfs}
\usepackage[title]{appendix}
\usepackage{xcolor}
\usepackage{color}
\usepackage{soul}
\usepackage{braket}
\usepackage{textcomp}
\usepackage{txfonts}
\usepackage{booktabs}
\usepackage{listings}
\usepackage{upgreek}
\usepackage[unicode=true,pdfusetitle,
 bookmarks=true,bookmarksnumbered=false,bookmarksopen=false,
 breaklinks=false,pdfborder={0 0 1},backref=false,colorlinks=true]
 {hyperref}
\usepackage{tikz}
\usepackage{xr}

\begin{document}
\title{\large \textbf{Witnessing non-stationary and non-Markovian environments with a quantum sensor}}

\author{John W. Rosenberg}
\affiliation{{Department of Chemical and Biological Physics and AMOS}, {Weizmann Institute of Science}, {Rehovot} {7610001}, {Israel}}

\author{{Mart\'{i}n} {Kuffer}}
\affiliation{{Centro\ At\textrm{\'{o}}mico Bariloche}, {CONICET, CNEA}, {S.\,C.\,de Bariloche} {8400}, {Argentina}}
\affiliation{{Instituto de Nanociencia y Nanotechnologia}, {CNEA, CONICET}, {S.\,C.\,de Bariloche} {8400}, {Argentina}}
\affiliation{{Instituto Balseiro}, {CNEA, Universidad Nacional de Cuyo}, {S.\,C.\,de Bariloche} {8400}, {Argentina}}

\author{{Inbar} {Zohar}}
\affiliation{{Department of Chemical and Biological Physics and AMOS}, {Weizmann Institute of Science}, {Rehovot} {7610001}, {Israel}}

\author{{Rainer} {St\"ohr}}
\affiliation{{3.\,Physikalisches Institut}, {Universität Stuttgart}, {Stuttgart} {70569}, {Germany}}

\author{{Andrej} {Denisenko}}
\affiliation{{3.\,Physikalisches Institut}, {Universität Stuttgart}, {Stuttgart} {70569}, {Germany}}

\author{{Analia} {Zwick}}
\affiliation{{Centro\ At\textrm{\'{o}}mico Bariloche}, {CONICET, CNEA}, {S.\,C.\,de Bariloche} {8400}, {Argentina}}
\affiliation{{Instituto de Nanociencia y Nanotechnologia}, {CNEA, CONICET}, {S.\,C.\,de Bariloche} {8400}, {Argentina}}
\affiliation{{Instituto Balseiro}, {CNEA, Universidad Nacional de Cuyo}, {S.\,C.\,de Bariloche} {8400}, {Argentina}}

\author{{Gonzalo A.} {\'{A}lvarez}}
\email{gonzalo.alvarez@conicet.gov.ar}
\affiliation{{Centro\ At\textrm{\'{o}}mico Bariloche}, {CONICET, CNEA}, {S.\,C.\,de Bariloche} {8400}, {Argentina}}
\affiliation{{Instituto de Nanociencia y Nanotechnologia}, {CNEA, CONICET}, {S.\,C.\,de Bariloche} {8400}, {Argentina}}
\affiliation{{Instituto Balseiro}, {CNEA, Universidad Nacional de Cuyo}, {S.\,C.\,de Bariloche} {8400}, {Argentina}}

\author{{Amit} {Finkler}}
\email{amit.finkler@weizmann.ac.il}
\affiliation{{Department of Chemical and Biological Physics and AMOS}, {Weizmann Institute of Science}, {Rehovot} {7610001}, {Israel}}

\begin{abstract}
{Quantum sensors offer exceptional sensitivity to nanoscale magnetic field fluctuations, where non-stationary effects—such as spin diffusion—and non-Markovian dynamics arising from coupling to few environmental degrees of freedom play critical roles. Because fully reconstructing the microscopic structure of realistic spin baths is often infeasible, a practical challenge is to identify the dynamical features that are actually encoded in the sensor’s decoherence signal. Here, we demonstrate how quantum sensors can operationally characterize the statistical nature of environmental noise, distinguishing between stationary and non-stationary behaviors, as well as Markovian and non-Markovian dynamics. Using nitrogen-vacancy (NV) centers in diamond as a platform, we develop a physical noise model that captures the essential dynamical features of realistic environments relevant to sensor observables—independently of the microscopic bath details— and provides analytical predictions for Ramsey decay across different regimes. These predictions are experimentally validated through controlled noise injection with tunable correlation properties. Our results showcase the capability of quantum sensors to isolate and identify key dynamical properties of complex environments, without requiring full microscopic bath reconstruction. This work clarifies the operational signatures of non-stationarity and non-Markovian behavior at the nanoscale and lays the foundation for strategies that mitigate decoherence while exploiting environmental dynamics for enhanced quantum sensing.}
    
\end{abstract}

\maketitle

\section{Introduction}\label{sec1}

Quantum technologies \cite{Acin2018, Awschalom2018, Deutsch2020} rely on preserving quantum coherence for both information processing and sensing. In quantum computing, decoherence from environmental noise limits fidelity and scalability \cite{Zurek2003, Khodjasteh2009, Taminiau2014, Zhang2015}, making noise characterization essential. In quantum sensing, by contrast, decoherence serves as a resource: the sensor’s sensitivity to its environment enables probing of physical systems via their noise signatures~\cite{Degen2017, Aslam2023}. Understanding decoherence is thus key to both mitigating its effects and harnessing it for precision sensing at the nanoscale.

Noise-induced decoherence is ubiquitous in quantum systems, appearing in contexts ranging from nuclear spin baths \cite{Maze2008, Zhao2012, Faribault2013} and hyperfine interactions in diamond \cite{Wang2013, Park2022} to quantum dots \cite{Hanson2007}, superconducting qubits~\cite{Simmonds2004} and silicon donor defects~\cite{Kane1998}. Techniques such as dynamical decoupling and quantum error correction have been developed to counteract decoherence \cite{Khodjasteh2005, Uhrig2007, Du2009, Souza2011, Suter2016}, but their effectiveness relies on detailed knowledge of the noise characteristics \cite{Alvarez2010, Clausen2010, Alvarez2011, BarGill2012, Malinowski2016, Bylander2011, Sung2019, Wise2021}. Most existing methods, moreover, assume that the noise is stationary—i.e., that its statistical properties do not change in time \cite{Clausen2010, Alvarez2011, Bylander2011, Zwick2016}. 

At the nanoscale, where quantum sensors such as atomic defects in solids operate \cite{Barry2020, Budakian2024}, the sensing radii are typically on the order of 10 nm or less—far below those of conventional magnetic resonance techniques \cite{Janitz2022, Yudilevich2022, Dwyer2022, Liu2022}. In this regime, environmental dynamics near the sensor, such as spin diffusion, often dominate and exhibit non-Markovian and non-stationary characteristics \cite{alvarez_localization-delocalization_2015,Li2024, Wang2021, Norambuena2020, Kairys2023, Penshin2024, Ginot2022,dominguez_decoherence_2021}. These features can arise when the environment retains memory of its past or is driven out of equilibrium by local perturbations, respectively. 
Non-stationarity may originate either intrinsically, from sensor back-action or local perturbations, or be externally imposed through driven out-of-equilibrium conditions. For example, the act of initializing a quantum sensor—such as optical initialization—can induce a quench, resetting local environmental degrees of freedom and triggering intrinsic non-stationary dynamics due to quantum back-action~\cite{Wang2021, Jerger2023}. In contrast, sudden temperature changes, chemical reactions, environmental excitations, or phase transitions can externally drive the bath away from equilibrium~\cite{polkovnikov_colloquium:_2011,alvarez_localization-delocalization_2015,Li2024,lewis-swan_dynamics_2019}.

While such behavior complicates efforts to mitigate decoherence in quantum information platforms, it also offers opportunities: the decoherence itself becomes a signal, encoding the statistical properties of the surrounding environment. Quantum sensors therefore offer a powerful platform for probing these complex dynamics. However, unlocking this potential requires systematic frameworks for identifying and distinguishing between different types of environmental noise. Despite their importance, non-stationary~\cite{Ban2007, Booker2020, Basit2020, Kuffer2022, Kuffer2025_PRXQuantum} and non-Markovian \cite{Wolf2008, Cui2008, Liu2011, Penshin2024} noise processes remain poorly understood and difficult to characterize, limiting progress in both quantum control and nanoscale sensing.

Despite their ubiquity and relevance, there remains no general framework for identifying or classifying the statistical properties of real noise environments at the nanoscale. Each environment—whether biological, solid-state, or chemical—has unique microscopic details, making first-principles modeling impractical in most cases. Nevertheless, many such systems share common features, such as memory effects \cite{Park2022,Kuffer2022} or departures from equilibrium \cite{Wang2013,lewis-swan_dynamics_2019,Wang2021,Jerger2023,Li2024,Guo2024,polkovnikov_colloquium:_2011,alvarez_localization-delocalization_2015,dominguez_decoherence_2021}. What is needed are experimentally validated strategies that can isolate and characterize these features independently of system-specific complexity.

While real environments—such as paramagnetic spin baths—are governed by complex quantum many-body dynamics, a quantum sensor interacts with them through effective stochastic magnetic fields. As a result, fully reconstructing the microscopic bath is not only a formidable challenge in realistic settings, but also often unnecessary for sensing tasks. Instead, what the sensor provides is partial but highly informative access to the statistical structure of the environment through its decoherence signal. A more practical and experimentally relevant path forward is to, rather than model every microscopic degree of freedom, first identify the dynamical features that are operationally encoded in the measured sensor coherence—such as whether the bath is stationary or out of equilibrium, and whether memory effects are present. Our approach embraces this perspective: rather than aiming for complete bath reconstruction, it focuses on capturing the dynamical properties that directly manifest in the sensor response, providing a classification aligned with experimentally accessible observables.

To this end, we introduce a general model that captures key aspects of realistic noise processes—namely, whether they are stationary or non-stationary, and Markovian or non-Markovian. The model is analytically tractable and allows us to derive closed-form predictions for the Ramsey decay of a quantum sensor coupled to such noise. Crucially, we demonstrate that these four dynamical regimes (stationary Markovian, stationary non-Markovian, quenched Markovian, and quenched non-Markovian) exhibit qualitatively distinct signatures in the short- and long-time behavior of the sensor signal. This enables us to classify the dynamical nature of an environment directly from the Ramsey response, without assuming prior knowledge of the full noise correlation functions.

The model reduces to the well-known Ornstein–Uhlenbeck (OU) process in the Markovian limit, but can capture non-Markovian statistical features essential to quantum sensing. As a Gaussian model, it can also be viewed as a particular case of a Caldeira-Legget (CL) bath. However, while CL baths may exhibit arbitrary spectral densities and are typically analyzed at equilibrium, our framework focuses on the physically motivated subset that remains analytically tractable yet rich enough to capture out-of-equilibrium and non-Markovian dynamics. This makes it ideally suited to classify stationary versus non-stationary and Markovian versus non-Markovian regimes in quantum sensing scenarios.

We validate these predictions using a single nitrogen-vacancy (NV) center in diamond as a quantum sensor. Rather than relying on an uncontrolled physical bath, we emulate the statistical structure of the model environments through injected noise with tunable parameters. This level of control allows us to isolate dynamical features ---the impact of memory, damping, and initial-state quenches on the Ramsey decay--- that would coexist in a real spin bath, making the approach a proof-of-principle demonstration of what a quantum sensor can, in practice, identify about its environment. We find strong agreement between experimental measurements and theoretical predictions, confirming that a quantum sensor can act as a witness of the environmental noise’s statistical nature.

By establishing a pathway to disentangle and identify non-Markovian and non-stationary features, our work opens new directions for nanoscale noise characterization. Beyond offering a robust foundation for quantum sensing in complex environments, these results also have direct implications for improving decoherence control in quantum technologies, where understanding the nature of noise is a prerequisite for its mitigation~\cite{Wolf2008,Cui2008,Liu2011,Ban2007,Booker2020,Basit2020,Kuffer2022,Kuffer2025_PRXQuantum}. We emphasize that the framework is designed to classify and identify dynamical regimes rather than to reconstruct the full microscopic structure of arbitrary baths—a task that generally requires additional complementary spectroscopy techniques ~\cite{Kuffer2022}. The present approach therefore provides a practical foundation for interpreting the decoherence signatures observed in realistic sensing scenarios.

\begin{figure*}[!ht]
    \centering
    \includegraphics[width=\textwidth]{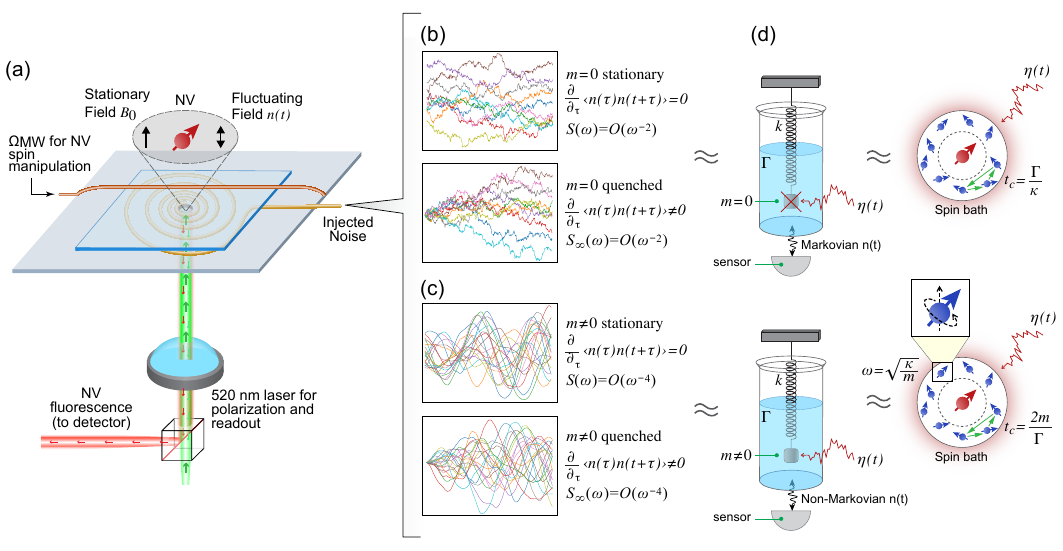}
    \caption{{Experimental setup to probe environmental noise using a single NV center as a quantum sensor via Ramsey decay measurements}. (a) The setup consists of a diamond membrane containing nanopillars with single NV centers placed on a printed circuit board. A 520\,nm laser is focused through a hole in the board to initialize and read out the NV spin state. The NV is subjected to a stationary magnetic field $B_0=298~\text{G}$, while environmental field fluctuations $n(t) = {\gamma}_\mathrm{nv}\Delta B_z\left(t\right)\ $  are generated by injecting noise through a spiral antenna beneath the diamond via a variable voltage. Radiofrequency (RF) pulses for NV spin manipulation are applied via a copper wire above the diamond membrane. (b,c) Representative samples (10 sample trajectories shown as colored curves for each) of the four noise processes $n(t)$ considered in this study: stationary Markovian, quenched Markovian, stationary non-Markovian, and quenched non-Markovian (from top to bottom). We distinguish between Markovian (panel b, $m = 0$) and non-Markovian (panel c, $m \neq 0$). (d) Schematic representation of how magnetic field fluctuations $n(t)$, sensed by the quantum sensor, are modeled as the position of a stochastically driven harmonic oscillator by a Langevin-type equation: $mn''+\Gamma n'+\kappa n=\eta(t)$. The parameters $m$, $\Gamma$, $\kappa$ describe the effective dynamics of the spins that directly couple to the qubit, capturing both oscillatory (non-Markovian) and purely relaxational (Markovian) regimes. The stochastic force $\eta(t)$, illustrated by the red arrow, represents the influence of more distant environmental spins that do not couple directly to the sensor but modulate the collective spin-bath dynamics. This model unifies  non-Markovian and Markovian spin-bath behaviors within a single physical framework.}
    \label{fig: setup}
\end{figure*}

\section{Analytical Framework for Quantum Sensing of Stationary and Non-Stationary, Markovian and Non-Markovian Noise}\label{am}
\subsection{The NV quantum sensor}
The NV quantum sensor is a spin-1 defect that detects the axial component (with respect to the N$\rightarrow$V axis) $B_z\left(t\right)=\ B_0+\ \Delta B_z\left(t\right)$ of a magnetic field through its Zeeman interaction (Fig.\,\ref{fig: setup}(a)). By considering a fluctuating field along the N$\rightarrow$V axis, we remain within the framework of pure dephasing and neglect other relaxation processes. This approximation is well justified whenever the total measurement duration is much shorter than the longitudinal relaxation time $T_1$ of the NV center. We use two spin substates, $\ket{0}$ and $\ket{-1}$, to form the qubit \cite{Doherty2013}. By preparing a superposition of these states, the relative phase accumulated between them over time reflects the magnetic field fluctuations. Applying microwave pulses at the qubit's Larmor frequency allows one to use the rotating wave approximation, where the stationary term ${\gamma}_\mathrm{nv}B_0t$ is canceled, and where ${\gamma}_\mathrm{nv}$ is the gyromagnetic ratio of the NV center. The accumulated phase is then given by $\varphi (t)=\int^t_0{{\gamma}_\mathrm{nv}\Delta B_z}\left(t'\right)dt'$, and the qubit state projection onto {\textbar}0$\mathrm{\rangle }$ will oscillate as ${\mathrm{cos} \left(\int^t_0{{\gamma}_\mathrm{nv}\Delta B_z\left(t'\right)}dt'\right)\ }$. For more details see Appendix \ref{sec:procedure_Ramsey_innermost_loop}.

When the environment induces pure dephasing and the noise is Gaussian, the semiclassical approximation becomes exact: the effect of the environment—even if quantum in nature—can be fully described by a classical stochastic process $n(t)$~\cite{Szankowski2017}. This applies to a broad class of physical systems, including spin baths and bosonic environments under weak coupling to the sensor \cite{Alvarez2011, BarGill2012, Bylander2011,Faribault2013,Wang2013, Park2022}. Modeling $n(t)$ as a Gaussian process is therefore both physically justified and analytically advantageous: It captures essential features of non-Markovian and non-stationary noise, as we show below, while avoiding the added complexity of higher-order cumulants. Gaussian noise also provides a paradigmatic and sufficient case for developing the classification framework. Moreover, the qualitative behaviors we identify—such as short-time scaling laws associated with the different regimes and oscillatory correlations signaling non-Markovianity—are expected to remain valid beyond the Gaussian limit, for instance in non-linear Langevin-type dynamics that generate non-Gaussian statistics. This provides a general and tractable framework for identifying signatures of complex environmental dynamics through quantum sensors.

We calculate the Ramsey signal $S(t)$ by averaging over all noise trajectories (see Fig.\,\ref{fig: setup}(b), (c)). Since $n(t)$ is a Gaussian process with mean zero, $\varphi (t)$ is also Gaussian with mean zero and variance ${\sigma }^2_t$. This leads to a signal decay of 
\begin{equation}
S(t)=\left\langle \cos(\varphi(t))\right\rangle =\int^{\infty }_{-\infty }{\frac{1}{{\sigma }_t\sqrt{2\pi }}e^{-\frac{{\varphi }^2}{2{{\sigma }_t}^2}}e^{i\varphi }d\varphi }=e^{-\frac{{{\sigma }_t}^2}{2}}
\label{eq: signal_decay_gaussian}
\end{equation}

\noindent The correlation function $\left\langle n(t_1)n(t_2)\right\rangle $ of the environmental noise evaluated at two different times $t_1$ and $t_2$, dictates the attenuation factor
\begin{equation}
\chi \left(t\right)=\frac{{{\sigma }_t}^2}{2}=\int^t_0{\int^{t_1}_0{\left\langle n\left(t_1\right)n\left(t_2\right)\right\rangle dt_2}dt_1},
\label{eq: attenuation_factor}
\end{equation}
providing an exact expression for the Ramsey decay $S(t)\ =\ e^{-\chi \left(t\right)}$. In real experiments, additional sources of dephasing or relaxation may superimpose an extra exponential envelope on the Ramsey signal. Our framework therefore establishes a baseline scenario that isolates the contribution of the target environment from other noise channels. The predicted signatures remain observable as long as the corresponding decay rates are not dominant over those of the environment under study.

\subsection{Environmental fluctuation model}
To capture essential features of non-Markovian and non-stationary environmental noise affecting the qubit, we introduce a general model for the fluctuating magnetic field using a semiclassical stochastic noise process $n(t)$. Rather than targeting a specific microscopic system, this approach aims to reproduce universal statistical behaviors—such as memory effects and out-of-equilibrium evolution—that are expected in realistic quantum environments at the nanoscale.

The temporal correlations of $n(t)$ are described by a Langevin-type equation 
\begin{equation}
mn''+\Gamma n'+\kappa n=\eta(t),
\label{eq: Langevin}
\end{equation}
where $m$ denotes an effective mass, quantifying the inertia of the fluctuating field; $\Gamma$ is a damping coefficient, and $\kappa$ corresponds to a potential confining the fluctuations (Fig.\,\ref{fig: setup}(d)). The stochastic driving term $\eta \left(t\right)$ represents Gaussian white noise process with zero mean and autocorrelation
\begin{equation}
\left\langle \eta \left(t_1\right)\eta \left(t_2\right)\right\rangle =A\delta (t_1-t_2),
\label{eq: gaussian_white_noise}
\end{equation}
capturing random forces with strength $A$ from remote environmental degrees of freedom that do not directly interact with the sensor. 

This model is physically grounded and versatile: as we show in Appendix \ref{sec:physical_realizations}, it captures the dynamics of diverse systems including a quantum harmonic oscillator coupled to thermal baths (e.g., Caldeira–Leggett-type models), and spin baths with varying internal couplings as shown in Fig.\,\ref{fig: setup}(d). The parameters $m$, $\Gamma$ and $\kappa$ encode the properties of environmental modes that interact with the qubit, capturing an interplay between oscillatory and relaxation dynamics. Note that $m$ may coincide with a physical mass, as in the damped harmonic oscillator case (Appendix \ref{subsec:Damped_oscillator}), or act as an effective parameter encapsulating multiple physical quantities that determine the inertia of the fluctuating field, as in a spin bath (Appendix \ref{subsec:Asymmetric_spin_bath}). The term $\eta(t)$ accounts for stochastic influences from remote or indirectly coupled environmental degrees of freedom.

A critical feature of this model is the \textit{inertia term}: for $m\neq0$, the environmental dynamics acquire memory, with future fluctuations depending on past velocities and positions. This generates time-correlated, non-Markovian noise with continuous derivatives \cite{Kuffer2022}. In the limit $m = 0$, the dynamics reduce to an overdamped OU process---fully memoryless and Markovian. When $m\neq0$, the inertia term introduces a finite memory time, enabling transitions from purely Markovian to non-Markovian behavior. This parameter thus governs the degree of temporal correlations in the noise.

In quantum sensing, the objective is to extract the dynamical properties of the environment itself from the sensor's response. This differs from approaches that focus on classifying the non-Markovianity of the sensor’s reduced dynamics~\cite{Haase2018}. Here, we distinguish whether the environmental fluctuations $n(t)$ are Markovian or non-Markovian—based on the presence ($m \ne 0$) or absence ($m = 0$) of memory in their dynamics.

The model also distinguishes between \textit{stationary} and \textit{non-stationary} noise based on initial conditions (Appendix \ref{sec:Derivation_correlation_functions}). Equilibrium states yield stationary noise with correlation functions invariant under time translation, while out-of-equilibrium initialization---such as from quantum back-action---induces non-stationary noise with time-dependent correlation functions. This is particularly relevant for quantum sensors, where initialization perturbs the environment \cite{Wang2021, Jerger2023}. When the noise is stationary, our model corresponds to a Gaussian limit that can be formally mapped to a Caldeira–Leggett (CL) bath with an appropriate spectral density. However, out-of-equilibrium initial conditions are naturally incorporated within our framework, whereas general CL formulations typically assume equilibrium environments and require additional assumptions to describe quenched or time-dependent states.

Together, the inertia parameter $m$ and the initial conditions of the bath define the four fundamental regimes analyzed in this work: stationary and non-stationary, Markovian and non-Markovian. Each regime produces distinct Ramsey decay signatures, enabling the quantum sensor to diagnose the underlying noise statistics. This framework thus provides a tractable and physically grounded basis for modeling noise-driven decoherence: it admits analytical treatment, enabling systematic investigation of how different environmental dynamics—and their interplay—manifest in the sensor's response.

We remark that our approach is agnostic to bath details, focusing on detecting nonstationary and non-Markovian behaviour. We use a simple model to capture essential environmental characteristics, while remaining analytically tractable and allowing us to isolate specific dynamical features (nonstationarity and non-Markovianity). This enables the detection of these specific features without the need of fully characterizing the noise spectrum/correlation function of the environment. Our results do not provide a full reconstruction of the environmental noise characteristics, like noise spectroscopy would do. Instead, our method uses less resources to extract key information from the environmental behavior. Extensions to fully quantum, non-Gaussian and other more complex environments are possible, but fall beyond the scope of this work.

\subsection{Noise correlation characteristics}
The correlation function of environmental noise, which defines the statistical behavior of  $n(t)$, determines the Ramsey decay and allows us to distinguish between different types of noise. We examine two limits of the unified stochastic model: Markovian ($m\to0$) and underdamped non-Markovian ($4\kappa m>\Gamma^2$).

For $m=0$, representing extremely overdamped systems, the noise behaves as constrained Brownian motion, described by $n'+t_c^{-1}n=\eta(t)/\Gamma$, where $t_c=\Gamma/\kappa $ is the self-correlation time of the environmental fluctuations for the $m=0$ case, which determines the characteristic timescale over which the environment reaches its equilibrium (stationary) state. This is Markovian noise, with a propagator that lacks memory of previous states. 

For~$m\neq 0$, which we use to represent underdamped systems, the noise behaves as a stochastically driven damped harmonic oscillator: $n''+2t^{-1}_c n'+\omega_0^2n=\eta(t)/m$, where $t_c = 2m/\Gamma$ is the self-correlation time for the $m\neq 0$ case, and $\omega_0 =\sqrt{\kappa/m}$ is the restoring frequency. This same equation also describes overdamped systems. In these cases, oscillations are suppressed and the self-correlation function decays bi-exponentially, as observed for nuclear spin baths under strong dephasing \cite{alvarez2006,Danieli2007}. Such dynamics still exhibit non-Markovian memory due to the coexistence of two distinct relaxation rates, which cannot be reproduced by a purely Markovian Ornstein–Uhlenbeck process. 

The inertia term ($m \neq 0$) introduces an additional timescale, $\omega_0$, such that the coexistence of $\omega_0$ and $t_c$ couples the dynamics to past velocities and positions, thereby generating non-Markovian memory effects. In the underdamped regime ($\omega_0 t_c > 1$), the resulting oscillatory correlations with an effective frequency $\Omega=\sqrt{\omega_0^2-t^{-2}_c}$ are a hallmark of many non-Markovian processes (derivations in Appendix \ref{sec:Underdamped_correlation_functions}).

This model effectively captures the behavior of spin baths with randomly interacting spins, where the correlation function exhibits a dynamical phase transition between overdamped and underdamped behavior depending on the structure of the environmental Hamiltonian. 
 Such dynamics are beyond the scope of Markovian descriptions \cite{Li2024}. It also applies to other relevant quantum environments, including thermally populated bosonic modes and quantum spin systems (Appendix \ref{sec:physical_realizations}).

For a nuclear spin bath, for example, $\omega_0$ corresponds to the Larmor frequency, and $t_c$ denotes the spin decoherence time, as depicted in Fig.\,\ref{fig: setup}(d). In our experiments, we focus on the case $\omega_0 t_c > 1$ of underdamped dynamics, where the noise simultaneously oscillates and relaxes towards equilibrium, with an effective frequency $\Omega=\sqrt{\omega_0^2-t^{-2}_c}$. 

Both Markovian and non-Markovian noises may exhibit stationary or non-stationary behavior, depending on the initial state. The latter arises when the environment is initialized out of equilibrium—either via an external quench due to driven out-of-equilibrium phenomena, or through intrinsic quantum back-action from sensor initialization \cite{Wang2021, Jerger2023}.

The quantum sensor's sensitivity to these noise characteristics enables it to act as a witness to non-stationary and/or non-Markovian behavior. Figure\,\ref{fig: setup}(b) illustrates differences in correlation function behavior for the primary noise types. For stationary noise, the correlation function is time translation-invariant, $\left\langle n(t_1)n(t_2)\right\rangle =\left\langle n(0)n(\Delta t)\right\rangle $,  where $\Delta t=t_2-t_1$. For non-stationary noise, $\left\langle n(t_1)n(t_2)\right\rangle$ depends on both $t_1$ and $t_2$.

For Markovian noise, the spectral density $S_\infty(\omega)\propto \frac{1}{1 + t_c^2 \omega^2}$, defined as the Fourier transform of $\mathop{\mathrm{lim}}_{t_1\to \infty }\left\langle n(t_1)n\left(t_1+\Delta t\right)\right\rangle $, has a Lorentzian shape, i.e., is the inverse of an even quadratic polynomial in $\omega$, with noise fluctuations having discontinuous time derivatives. In non-Markovian noise ($m\neq 0$), $S_{\infty }(\omega )\propto \frac{ 1 }{ (\omega^2 - \omega_0^2)^2 + 4 t_c^{-2} \omega^2 }$ becomes the inverse of an even quartic polynomial, with continuity in noise derivatives. For our non-Markovian measurements, we focus on an underdamped regime where $S_{\infty }\left(\omega \right)$ peaks at $\pm\sqrt{\Omega^2-t_c^{-2}}$ (or at $0$ if $\Omega t_c< 1$). Detailed derivations of the correlation function $\left\langle n\left(t_1\right)n\left(t_2\right)\right\rangle$ and the Ramsey decay attenuation factor $\chi \left(t\right)$ are in Appendices \ref{sec:Derivation_correlation_functions} and \ref{sec:derivation_Ramsey_exponents}, respectively, for all the considered noises, as well as derivations of $S_{\infty }\left(\omega \right)$ for Markovian and underdamped noise in Appendix \ref{sec:Spectral_densities}.

\subsection{Ramsey decay for stationary and non-stationary Markovian noise}

The differences between stationary and non-stationary Markovian noise are evident in the short-time behavior of the Ramsey decay $\chi \left(t\right)$. For stationary (equilibrium) Markovian noise, the leading order term in the Ramsey decay is $\chi \left(t\right)=\frac{{\Delta}^2}{2}t^2+O(t^3)$. Note that it is proportional to $t^2$ and depends only on the equilibrium variance ${\Delta }^2={\mathop{\mathrm{lim}}_{t\to \infty } \left\langle {n(t)}^2\right\rangle \ }$, and not on the correlation time  $t_c$. In contrast, for quenched (non-stationary, with an initial condition~$n_0=0$) Markovian noise, the attenuation factor is $\chi \left(t\right)\approx \frac{{\Delta }^2}{3t_c}t^3+O(t^4)$, depending on both ${t_c}$ and ${\Delta }^2$. Alternatively, the attenuation factor can also be written as $\chi \left(t\right)\approx \frac{2}{3}\frac{A}{{\Gamma}^2}t^3+O\left(t^4\right)$, where we can see that the leading order of the decay factor of the quenched Markovian noise depends only on the normalized driving strength for the Markovian process $\frac{A}{{\Gamma}^2}$. For further detail see Appendix \ref{sec:Markovian_noise_Ramsey_exponents}.

This indicates that, for short times and sufficiently large values of~$t_c$, the ratio between the Ramsey signals for quenched versus equilibrium Markovian noise can become significantly large, allowing for clear differentiation between the two. Additionally, the Ramsey signal can serve as a witness of various non-stationary noise effects, such as identifying the moment of a quench relative to the start time of the Ramsey measurement or detecting abrupt changes in correlation time.  

In the short-time regime, quenched noise induces a slower Ramsey decay—of higher-order in time—compared to equilibrium noise, underscoring the potential of Ramsey interferometry to distinguish between different noise types. In practice, the short-time scaling can be resolved provided that the bath correlation time exceeds the control pulse duration---a condition typically satisfied in NV-center experiments. 

At long times, both stationary and non-stationary noise—whether Markovian ($m=0$) or non-Markovian ($m\neq~0$)—exhibit attenuation factors that converge to the same decay rate. This shows that long measurement times are less efficient at detecting early-time non-stationary dynamics than measurement times $\lesssim t_c$. However, the curves remain offset by a constant shift, reflecting residual information from the initial non-stationary state (see Appendix \ref{sec:Markovian_noise_Ramsey_exponents} for derivation and Table \ref{tab:S2} in Appendix \ref{sec:summary_Ramsey_exponents}). This demonstrates that memory of the early-time dynamics is retained even after the environment has relaxed to stationarity.

These distinct short- and long-time dependencies also imply that the parameters $t_c$, $A/\Gamma^2$, 
and $\Delta^2$ affect different measurable aspects of the Ramsey response. 
At short times, the decay depends only on $\Delta^2$ for equilibrium noise and on $A/\Gamma^2$ for quenched noise, 
while $t_c$ governs the crossover between these regimes and the long-time offset. 
Consequently, a single measurement restricted to one regime cannot determine all parameters independently. 
However, by combining stationary and quenched experiments—which probe complementary temporal windows—it becomes 
possible to resolve these parameters unambiguously. 
This principle is analyzed quantitatively in Appendix~\ref{sec:Dependency_fitting_parameters}.

\subsection{Ramsey decay as a witness of non-Markovian noise: stationary vs.\,non-stationary}

The signatures of non-Markovian noise sources in Ramsey decay reveal a richer array of phenomena compared to Markovian counterparts. Here, we focus on underdamped non-Markovian ($m\neq 0$) noise. Sample plots of the analytical expressions for Ramsey decay curves  for equilibrium and quenched underdamped non-Markovian noise are shown in Fig. \ref{fig: analytical} in Appendix \ref{sec:Analytical_plots_nonMarkovian_noise}. 

For equilibrium (stationary noise), the leading order term of the Ramsey attenuation factor (see Appendix \ref{sec:underdamped_Ramsey_exponents} for derivation and Table \ref{tab:S2} in Appendix \ref{sec:summary_Ramsey_exponents}) is $\chi \left(t\right)\approx \frac{{\Delta}^2}{2}t^2+O(t^3)$. Note that the leading order coincides with the result for the stationary Markovian case. As before, the result depends only on the equilibrium variance ${\Delta}^2$ and is of order $t^2$. Note as well how this means that short-time ($t<{\Omega}^{-1},\ t_c$) Ramsey spectroscopy cannot distinguish between Markovian and non-Markovian noises when measuring them in a stationary state. This is intuitive, since (non-)Markovianity is a dynamical property, and thus the differences it induces can only be sensed after enough time has passed for the state of the noise to evolve. This result also holds in the overdamped regime, where the absence of oscillations does not alter the leading short-time $t^2$ scaling of $\chi(t)$.

In contrast, for quenched (non-stationary, with initial conditions $n(0)=n'(0)=0$) underdamped non-Markovian ($m \neq 0$) noise, the leading order of the attenuation factor is $\chi \left(t\right)\approx {\Delta}^2\frac{{\omega_0 }^2}{10t_c}t^5+O\left(t^6\right)$. Here, the behavior changes significantly: it is of order $t^5$, and depends not only on the equilibrium variance ${\Delta}^2$, but also on the self-correlation time $t_c$ and the oscillation frequency $\Omega$. This marks a clear departure from the quenched Markovian case. Thus, one can infer that short time Ramsey experiments can indeed distinguish between Markovian from non-Markovian environments, provided the proper quench is induced in these environments before the measurement. For further detail and derivation see Appendix \ref{sec:underdamped_Ramsey_exponents} and Table \ref{tab:S2} in Appendix \ref{sec:summary_Ramsey_exponents}.

This distinction is particularly relevant because it allows the ratio of Ramsey signals (quenched vs.\,equilibrium) under underdamped non-Markovian noise to become arbitrarily large at short times ($t<{\Omega}^{-1},\ t_c$), given sufficiently large $t_c$ and small $\omega_0$. This suggests that short-time Ramsey measurements can effectively discriminate between stationary and non-stationary non-Markovian noise. 

Finally, the attenuation factor can be expanded at short times as $\chi \left(t\right)\approx \frac{A}{40m^2}t^5+O\left(t^6\right)$, where the $t^5$ term represents the leading-order contribution for quenched non-Markovian noise with initial conditions $n(0)=n'(0)=0$. This leading order coefficient depends only on the normalized driving strength $\frac{A}{m^2}$, while the influence of other bath parameters, such as $t_c$ and $\omega_0$, appears only in higher-order terms ($\mathcal{O}(t^6)$ and beyond). The suppression of lower-order contributions is a direct consequence of the quench, which resets the bath away from equilibrium. The full expression for $\chi \left(t\right)$, including all parameter dependencies, is derived in Appendix~\ref{sec:underdamped_Ramsey_exponents}. We note that the same leading-order independence from other bath parameters also holds in the overdamped regime.

For both non-Markovian and Markovian noise, reducing the damping (i.e., increasing $t_c$) causes the quenched Ramsey decay curve to slow down at short times ($t\ll t_c$) relative to the equilibrium Ramsey decay curve. However, the oscillatory nature of $\chi \left(t\right)$ for underdamped non-Markovian noise introduces a striking difference between the Ramsey signals for quenched vs.\,equilibrium noise. Specifically, for equilibrium underdamped non-Markovian noise with large $t_c$, the Ramsey signal exhibits collapses and revivals as shown in Fig.\,\ref{fig: analytical} of Appendix \ref{sec:Analytical_plots_nonMarkovian_noise}: the signal decays nearly completely, only to revive periodically, forming peaks at $t=\frac{2\pi m}{\Omega}$ for $m\in {\mathbb{Z}}^*$ (${\mathbb{Z}}^*$ denoting the positive integers), where $\Omega = \sqrt{{\omega_0 }^2-t^{-2}_c}$ is the effective oscillation frequency of the environment.  

In contrast, quenched underdamped non-Markovian noise does not produce the collapses and revivals seen in the stationary case (Fig.\,\ref{fig: analytical} in Appendix \ref{sec:Analytical_plots_nonMarkovian_noise}). Instead, the Ramsey signal exhibits plateaus at $t=\frac{2\pi m}{\Omega}$ for $m\in {\mathbb{Z}}^*$, with the steepest decay near $t=\frac{\pi (2m+1)}{\Omega}$. Despite the underlying memory, non-stationarity suppresses coherent revivals, replacing them with delayed decoherence and flat intervals—revealing how initial conditions critically shape non-Markovian dynamics.

These contrasting behaviors also illustrate how different dynamical features of the Ramsey response encode 
distinct environmental parameters. The oscillation frequency $\omega_0$ determines the revival period, 
the correlation time $t_c$ sets the envelope and plateau widths, and the overall noise strength $A/m^2$ 
controls the decay rate at short times. 
Consequently, a single stationary or quenched experiment samples only a subset of this information, 
and fitting all parameters independently may become ambiguous if the accessible timescales are limited. 
By combining stationary and quenched measurements---each sensitive to different physical features---one gains 
access to independent observables that together allow $\omega_0$, $t_c$, and $A/m^2$ to be extracted 
unambiguously. 
This principle is discussed in more detail in Appendix~\ref{sec:Dependency_fitting_parameters}.

The origin of collapses and revivals for equilibrium underdamped non-Markovian noise lies in a noise variance that is constant in time and the sharp peak in its noise spectrum near the oscillation frequency, at $\pm\sqrt{\Omega^2-t_c^{-2}}$ . This peak induces periodic cancellations of noise correlations and anticorrelations, producing revivals in the Ramsey signal at integer multiples of $\frac{2\pi }{\Omega}$. Such behavior cannot be seen with Markovian noise sources, where the correlation function is strictly non-negative, thus precluding the possibility of~collapses and revivals. Therefore, the presence of these revivals in the equilibrium Ramsey signal serves as a robust witness of non-Markovianity in the noise source.

Additionally, this model indicates another way in which Ramsey measurements can reveal out-of-equilibrium dynamics: As the starting time of the Ramsey measurement approaches a quench, the curvature of the Ramsey signal at revival points approaches zero, providing a distinct indicator of the transition to non-stationary noise. Thus, Ramsey measurements can serve as powerful tools for identifying both non-Markovian behavior and of out-of-equilibrium dynamics in complex noise environments.

\section{Experimental Demonstration of Quantum Sensing of Nonstationary and Non-Markovian Noise}

\subsection{Experimental implementation}
To test our theoretical predictions, we perform Ramsey measurements on a single NV center under injected noise with tunable correlation functions. Details of the noise generation algorithms and their validation against target correlation functions are provided in Appendices \ref{sec: methods} and \ref{sec:validation_injected_correlations}. While real environments are complex and system-specific, synthetic noise offers a controlled platform to isolate and probe key statistical features—such as non-stationarity and memory—relevant to nanoscale quantum sensing.

We use a diamond pillar containing a single NV center, with a sufficiently long $T_2^*$ time to ensure that the injected noise dominates the sensor’s decoherence. We manipulate the state of the NV using RF pulses generated by a wire passing above the diamond wafer, and we generate the injected noise by applying a voltage with the desired correlation function to a spiral antenna located beneath the diamond (see Fig.\,\ref{fig: setup}(a)). We use a confocal fluorescence microscope to focus a 520 nm laser on the NV for initialization and readout. For each time point in each Ramsey decay curve, we collect photons for 100 realizations of the injected noise, so that the projections of the NV state onto $|0\rangle$ are sensitive to the full statistical distribution of the noise realizations (see Appendix \ref{sec: methods}).

\subsection{NV characterization}
ODMR, Rabi, and Ramsey measurements for this NV are shown in Fig.\,\ref{fig: ramsey_char} of Appendix \ref{sec:NV_characterization}. To account for imperfections in the drift correction, we also carry out drift-corrected Ramsey measurements with 0\,V injected noise as a control and compare the result with a drift-corrected Ramsey measurement where a constant 0.45\,V is applied to the spiral. The results of these control measurements are also shown in Appendix \ref{sec:NV_characterization}. The Ramsey signal from the latter experiment shows an oscillation at 3.6\,MHz, which is significantly faster than the decay rate of 0.608 $\upmu$s${}^{-1}$, indicating that this NV is suitable for sensing the injected noise.

\subsection{Probing stationary versus non-stationary Markovian noise}
\begin{figure}[!ht]
    \centering
    \includegraphics[width=\columnwidth]{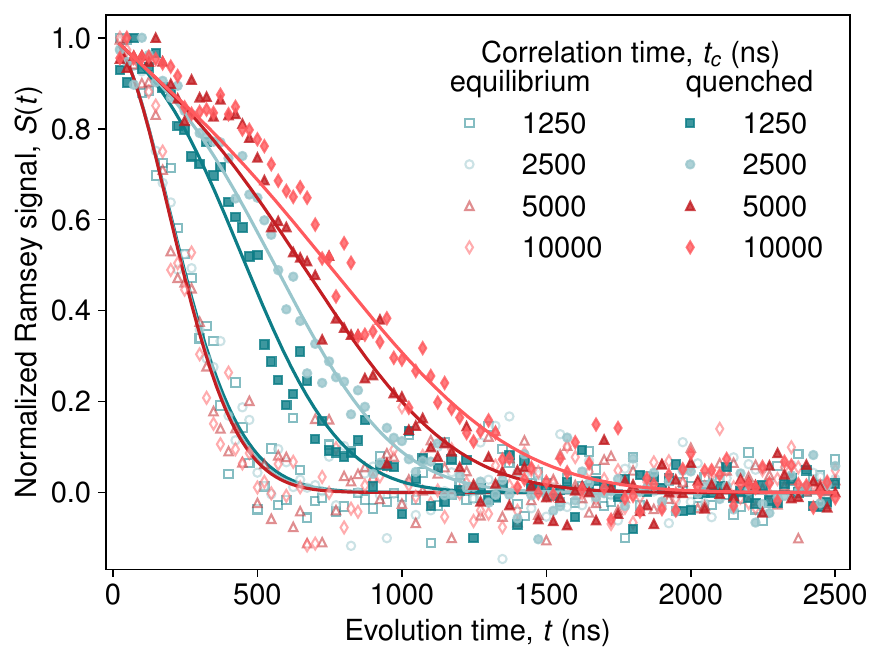}
    \caption{{Ramsey decay signals, ${S(t)}$, for equilibrium and quenched Markovian noise}, with correlation times \(t_c = 1.25, 2.5, 5,\) and \(10\,\upmu\mathrm{s}\), and noise standard deviation \(\Delta = 0.10\,\mathrm{V}\). Symbols represent experimental data; solid curves are analytical fits using \(t_c\) as a free parameter, consistent with the injected values. For equilibrium noise, all curves collapse onto a single decay due to identical short-time behavior, making them visually indistinguishable. In contrast, quenched noise curves remain clearly distinct, demonstrating the Ramsey signal’s sensitivity to the correlation time in non-stationary regimes.}
    \label{fig: ramsey_eq_non_eq}
\end{figure}

In the first set of injected noise measurements, we examine the effect of equilibrium Markovian and quenched Markovian noise with a standard deviation of $\Delta=$ 0.10\,V at equilibrium and correlation times $t_c=$ 10, 5, 2.5 and 1.25 $\upmu$s. Figure\,\ref{fig: ramsey_eq_non_eq} shows the Ramsey decays, along with fits using the analytical expressions for Ramsey decay for equilibrium and quenched Markovian noise shown in Eqs.\,(\ref{eq:S15}) and (\ref{eq:S17}).

\noindent One can see that the analytical expressions for equilibrium and quenched Markovian noise provide excellent fits for the experimental data. 

Notably, our results follow the short time-behavior predicted by the model, and they demonstrate that information about the correlation time can be extracted from a quenched environment---but not from an equilibrium environment---if the coupling with the environment is strong enough that stationary noise causes the Ramsey signal to decay significantly on a shorter timescale than the correlation time. 

The contrast between equilibrium and quenched Markovian noise is most evident at short evolution times. For equilibrium noise, the Ramsey signal decays significantly at times much shorter than the correlation time \( t_c \) and then reaches the noise floor, such that the decay curves for  different \( t_c \) values overlap. In contrast, in a quenched environment, the correlation time can be inferred before the signal significantly decays, improving the signal-to-noise ratio (SNR) for estimating the correlation time. Figure \ref{fig: ramsey_eq_non_eq} illustrates how the Ramsey decay curves for Markovian noise with different \( t_c \) remains distinct in the quenched case. Details of the fitting procedure used to extract the parameters $\Delta$ and $t_c$ are provided in Appendix \ref{sec:Dependency_fitting_parameters}. It is important to note that when a measurement probes only a restricted dynamical regime, 
the fitting parameters may become correlated, leading to over-parametrization. 
This is a universal feature of time-domain spectroscopy rather than a limitation of our model: 
if the measurement window does not encompass the relevant temporal scales, not all environmental 
parameters can be independently extracted. 
In our case, the equilibrium Ramsey experiment is mainly sensitive to the stationary variance $\Delta^2$, 
while the quenched Ramsey experiment isolates the driving strength $A/\Gamma^2$. 
By jointly analyzing both, one gains access to independent observables that together determine 
all relevant parameters of the noise, thereby resolving the apparent parameter correlations. 
This principle is analyzed in Appendix~\ref{sec:Dependency_fitting_parameters}.

\begin{figure*}[!ht]
\centering
\includegraphics[width=\textwidth]{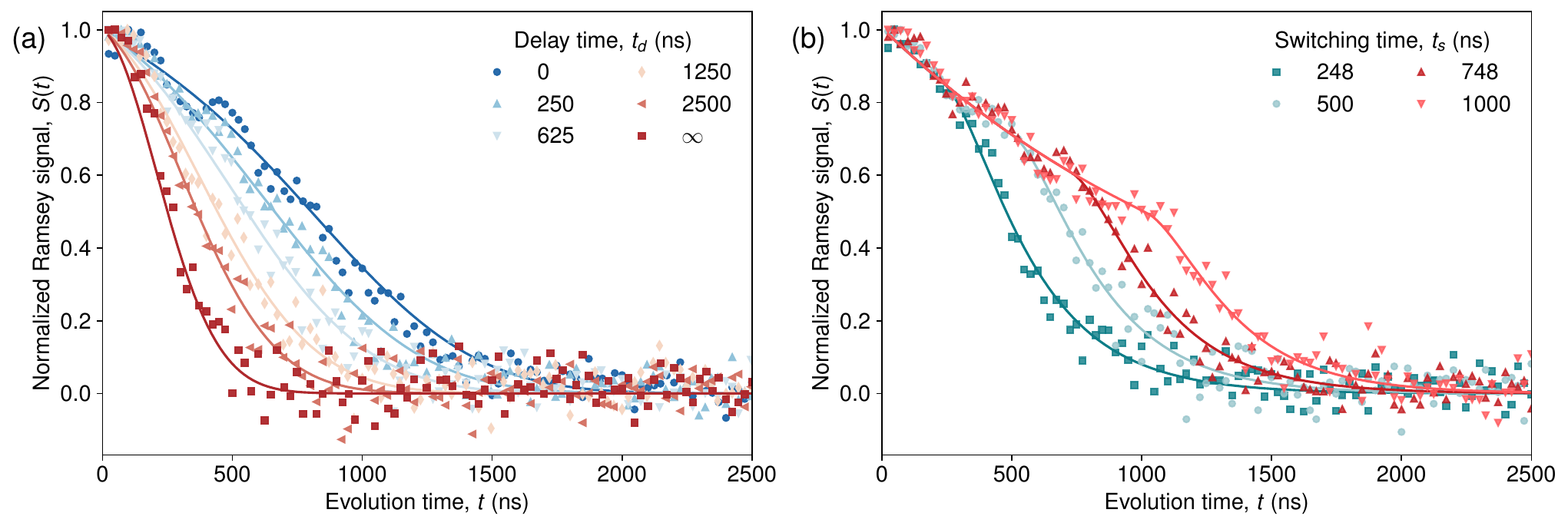}
\caption{{Measured Ramsey signals, ${S(t)}$, used to probe the onset of quenches in Markovian noise environments.} Symbols represent experimental data and solid lines are fits to the analytical expressions. (a) Ramsey decays for noise with a standard deviation $\Delta$ = 0.10\,V and correlation time $t_c = 10\,\upmu\mathrm{s}$ measured at equilibrium ($t_d = \infty$) and with quenches occurring at delays $t_d$ = 0, 250\,ns, 625\,ns, 1250\,ns, and 2500\,ns prior to the start of the Ramsey sequence. (b) Ramsey decays for noise with fixed  standard deviation $\Delta$ = 0.10\,V and a correlation time that switches from $t_a$ = 15\,ns to $t_b$ = 150\,ns at times $t_s$ = 248\,ns, 500\,ns, 748\,ns, and 1000\,ns.}
\label{fig: switch_delay}
\end{figure*}

To further demonstrate the utility of quantum sensors in extracting dynamical parameters from out-of-equilibrium environments, we investigated whether the timing of a quench can be inferred from the sensor response. Specifically, we measured Ramsey signals for both quenched and equilibrium Markovian noise with a standard deviation $\Delta=$ 0.10\,V at equilibrium and a correlation time $t_c=$ 10 $\upmu$s, and then introduced controlled delays $t_d$ of 250, 625, 1250, and 2500 ns between the quench and the start of the Ramsey measurement. The resulting signals are shown in Fig.\,\ref{fig: switch_delay}(a), along with fits using the analytical expression for the Ramsey signal of Markovian noise with a delayed quench derived in Eq.\,(\ref{eq:S19}), with the quench parameter $Q=e^{\frac{-2t_d}{t_c}}$ as a free parameter in the fit. The delay times $t_d$ extracted from the fits are shown in Table\,\ref{table: table1}, and match the known delay times used to generate the injected noise. These results confirm that, when the correlation time is known, the NV sensor’s Ramsey response can be used to determine when a quench occurred.

The Ramsey signal from an NV can also be used to determine when an abrupt switch in correlation time occurred relative to the start of the measurement. In a physical system, such a switch could arise, for example, from an abrupt change in temperature that changes the rate of molecular tumbling. To demonstrate this, we measured Ramsey signals for equilibrium Markovian noise with a constant standard deviation~$\Delta = 0.10$\,V and a correlation time that switches from $t_c=t_a$ = 15\,ns to $t_c=t_b$ = 150\,ns at switching times $t_s$ = 248\,ns, 500\,ns, 748\,ns, and 1000\,ns after the start of the measurement.
\noindent 
\begin{table}[!ht]
\begin{tabular}{|p{1.25in}|p{0.4in}|p{0.4in}|p{0.45in}|p{0.52in}|}
\hline 
Injected delay time, $t_d$ (ns) & $250$ & $625$ & $1250$ & $2500$ \\ \hline 
Fitted delay time, $t_d$ (ns) & $271 \mathrm{\pm} 30$  & $641 \mathrm{\pm} 48$ & $1418 \mathrm{\pm} 96$ & $2662 \mathrm{\pm} 198$ \\ \hline 
Injected switch time, $t_s$ (ns) & $248$ & $500$ & $748$ & $1000$ \\ \hline 
Fitted switch time, $t_s$(ns) & $235 \mathrm{\pm} 13$ & $504 \mathrm{\pm} 15$ & $728 \mathrm{\pm} 17$ & $1017 \mathrm{\pm} 20$ \\ \hline 
\end{tabular}
\caption{Comparison of the delay times $t_d$ and switch times $t_s$ extracted from fitting the Ramsey signals in Fig.\,\ref{fig: switch_delay} with the analytical expressions vs.\,the known values of the delay times and switch times for the injected noise.}
\label{table: table1}
\end{table}
The results are shown in Fig.\,\ref{fig: switch_delay}(b), along with fits using the analytical expression for the Ramsey signal under equilibrium Markovian noise with a switched correlation time, shown in the Appendix\,\ref{sec:Markovian_noise_Ramsey_exponents} (Eqs.\,(\ref{eq:S21}), (\ref{eq:S22})). The time of the switch relative to the start of the measurement was used as a free-fitting parameter. The switching times extracted from the fits are summarized in Table\,\ref{table: table1}, and show good  agreement with the known switching times used to generate the injected noise. Another example of switching time detection can be found in Appendix \ref{sec:additional_example}.

\subsection{Probing stationary versus non-stationary non-Markovian noise}
\begin{figure*}[!ht]
    \centering
    \includegraphics[width=\textwidth]{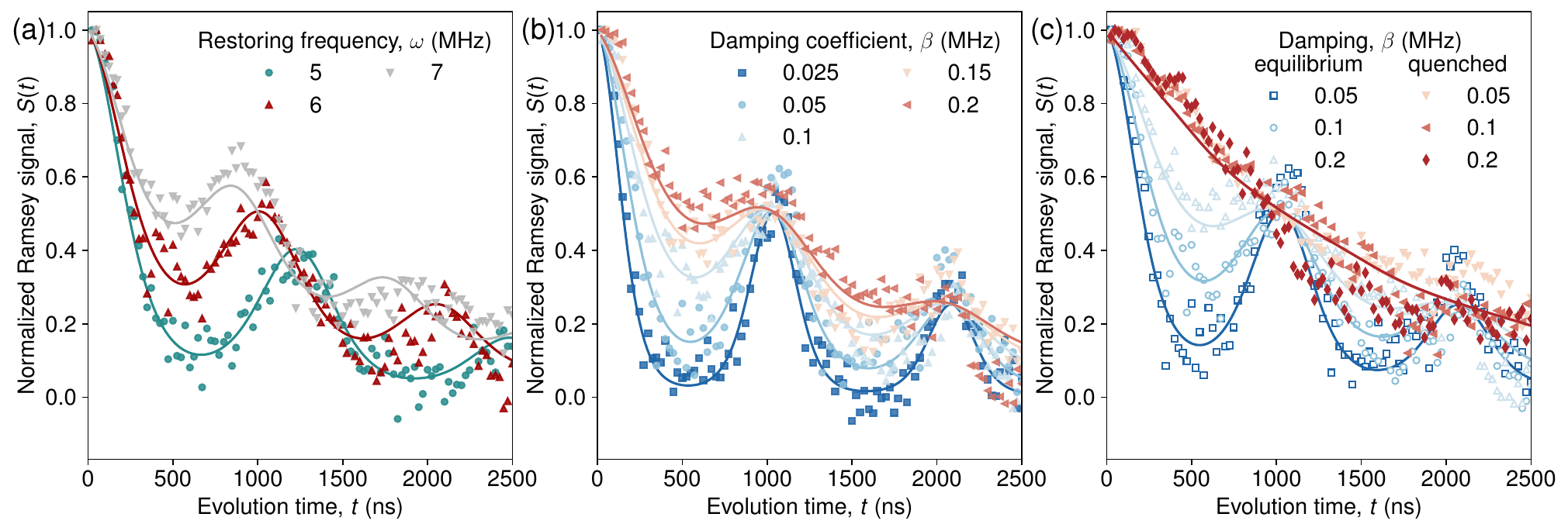}
    \caption{{Measured Ramsey decay signals ${S(t)}$ under equilibrium and quenched underdamped non-Markovian noise ($m \neq 0$), used to probe non-Markovian features and their interplay with non-stationary effects.} The  effective driving strength is fixed at $A/m^2$ = 0.054\,V$^2$/$\upmu$s$^3$. (a) Ramsey signals for a fixed damping coefficient $2t_c^{-1}$ = 0.1\,MHz, and restoring frequencies $\omega_0$ = 5, 6, and 7\,MHz. (b) Ramsey signals for a fixed restoring frequency $\omega_0$ = 6\,MHz and damping coefficients $2t_c^{-1}$ = 0.025, 0.05, 0.1, 0.15, and 0.2\,MHz. (c) Comparison of equilibrium and quenched non-Markovian noise for $\omega_0$ = 6\,MHz and $2t_c^{-1}$ = 0.05, 0.1, and 0.2\,MHz. In all cases, solid curves are fits to analytical expressions for Ramsey decay under non-Markovian noise (Appendix \ref{sec:underdamped_Ramsey_exponents}), using the experimental values of $\omega_0$ and $2t_c^{-1}$, with the effective driving strength of the magnetic field as the only free parameter to account for the antenna's impedance.}
    \label{fig: nm}
\end{figure*}

Lastly, we examine the effect of injected underdamped equilibrium non-Markovian ($m\neq 0$) noise. Figure~\ref{fig: nm}(a) shows the Ramsey signals measured upon injection of equilibrium non-Markovian noise with an effective driving strength $A/m^2\ $ of 0.054 V${}^{2}$/$\upmu$s$^{3}$, a damping coefficient $2{t_c}^{-1}$ of 0.1 MHz, and restoring frequencies $\omega_0$ = 5, 6, and 7\,MHz, along with fits using the analytical expressions for Ramsey decay under underdamped non-Markovian noise shown in Eqs.\,(\ref{eq:S24}), (\ref{eq:S26}). As expected, the observed Ramsey decay curves oscillate with an effective frequency $\Omega = \sqrt{\omega_0^2 - t_c^{-2}}$, so increasing the restoring frequency leads to earlier occurrence of  the first revival.

Figure~\ref{fig: nm}(b) shows Ramsey signals measured upon injection of equilibrium non-Markovian noise with the same effective driving strength $A/m^2$ of 0.054\,V${}^{2}$/$\upmu$s$^{3}$, a restoring frequency $\omega_0=6$ MHz, and varying damping coefficients $2{t_c}^{-1}$ of 0.025, 0.05, 0.1, 0.15, and 0.2\,MHz, along with corresponding analytical fits. For comparison, these damping coefficients are similar to values of ${T_2}^{-1}$ obtained for the P1 spin bath in nitrogen-doped diamond \cite{Park2022}. As expected, increasing  the damping coefficient reduces the amplitude of the collapse and revival features and broadens the revival peak. Importantly, if $\omega_0$ represents a bare precession frequency---such as the Larmor frequency of spins in a bath---the observed modulation would occurs at the effective frequency $\Omega = \sqrt{\omega_0^2 - t_c^{-2}}$, which deviates from the intrinsic value due to damping. This effect becomes evident only in the strongly damped cases.

The observation of collapses and revivals in the Ramsey signal—predicted by our analytical model for equilibrium underdamped non-Markovian noise—demonstrates that NV-based sensors can serve as witnesses of non-Markovian dynamics. These nontrivial features reflect how memory and dissipation reshape environmental fluctuations, enabling the sensor to distinguish between noise types and to access both bare and effective parameters. This highlights the relevance of the framework for extracting meaningful statistical information in realistic quantum sensing settings.

We also examined the effect of injected underdamped quenched non-Markovian noise with initial conditions $n(0)=n'(0)=0$. Figure~\ref{fig: nm}(c) shows Ramsey signals measured upon injection of quenched non-Markovian noise with an effective driving strength $A/m^2$ of 0.054\,V${}^{2}$/$\upmu$s${}^{3}$, a restoring frequency $\omega_0$ of 6\,MHz, and damping coefficients $2{t_c}^{-1}$ of 0.05, 0.1, and 0.2\,MHz, along with corresponding analytical  fits (solid lines). As predicted by the model (Appendices \ref{sec:underdamped_Ramsey_exponents} and \ref{sec:Analytical_plots_nonMarkovian_noise}), the collapses and revivals features are absent under quenched noise, offering a clear witness of a quench dynamics in underdamped non-Markovian  environments. This highlights the importance of careful analysis to ensure that key non-Markovian characteristics are not overlooked.

Moreover, note that all curves corresponding to the quenched cases in Fig.\,\ref{fig: nm}(c) overlap. This is because the Ramsey signal remains above the noise floor only for times much shorter than the correlation time $t_c$, where the leading-order term $\frac{A}{40m^2}t^5$ in the attenuation factor dominates. As a result,  the attenuation factor is solely determined by the normalized driving strength $\frac{A}{m^2}$, which is held constant in these experiments. 

Overall, these results demonstrate that the Ramsey signal of an NV center can serve as a sensitive probe of both out-of-equilibrium dynamics—such as quenches—and non-Markovian noise. Crucially, they underscore the importance of accurately resolving the interplay between memory effects and non-stationarity to avoid mischaracterizing the environment in realistic quantum sensing scenarios.

\section{Conclusions}
Using quantum sensors as witnesses of the statistical nature of noise dynamics offers several advantages. The agreement between the measured Ramsey decay curves and the predictions of our analytical model demonstrates the utility of such models in characterizing noise sources with non-stationary and non-Markovian dynamics—two key features that remain difficult to identify in realistic quantum environments. Having demonstrated that the decay attenuation factor for short evolution times is of higher order for quenched noise, we establish that Ramsey measurements can act as sensitive witnesses of non-stationary noise dynamics. Even when experimental resolution limits the extraction of short-time exponents, complementary signatures at longer times--such as offsets in $\chi(t)$ or the suppression of revivals--provide robust alternative indicators of non-stationary and non-Markovian behavior. This enables us to assess whether a noise source deviates from equilibrium under specific experimental conditions, either from intrinsic non-stationarity caused by sensor back-action or initialization, or from externally imposed non-stationarity due to driven dynamics. Furthermore, the observation of collapses and revivals in Ramsey decay for stationary underdamped non-Markovian noise highlights the potential of Ramsey measurements to reveal non-Markovian dynamics in noise sources. The results show that the sensor signal encodes not only decoherence rates, but also the underlying statistical nature and the nontrivial interplay between non-stationarity and non-Markovianity in the environment. While other dephasing or relaxation channels can obscure the signatures of non-Markovian and non-stationary noise, our experiments demonstrate that the proposed protocol remains viable provided these channels are not significantly stronger than the target environment. In this sense, our framework establishes a baseline scenario that isolates the contribution of environmental noise statistics, offering a clear foundation for interpreting quantum-sensing experiments.

The proposed framework not only classifies dynamical noise but also identifies which regimes must be accessed to extract all environmental parameters unambiguously, and how quenching can recover information otherwise inaccessible in stationary measurements. This operational perspective addresses a key challenge in nanoscale noise spectroscopy: full microscopic reconstruction is often impossible, but many practically relevant dynamical features—such as memory, departures from equilibrium, or the presence of few-body dynamics—can be inferred directly from the measured sensor response. Our method therefore complements, rather than replaces, full noise-spectroscopy approaches by providing a lightweight and experimentally accessible route to detect and interpret nonstationarity and non-Markovianity without requiring complete reconstruction of the correlation function.

Characterizing such dynamical features grants insights into the underlying physical mechanisms of  the noise—such as hyperpolarization, spin diffusion or back-action-induced quenches—and also lays the groundwork for developing strategies to mitigate or control these effects. Such strategies may involve optimizing experimental conditions to influence the noise bath or designing pulse sequences tailored to suppress specific noise types, particularly by accounting for memory timescales and deviations from equilibrium \cite{Alvarez2010, Clausen2010, Alvarez2011, BarGill2012, Malinowski2016, Bylander2011, Sung2019, Wise2021}. These approaches are essential for enhancing the performance and reliability of quantum technologies.

 While the present work focuses on Ramsey decay measurements, it opens a pathway to more comprehensive sensing protocols. Future studies could further enhance the robustness of noise characterization by incorporating additional pulse sequences, such as a Hahn echo, to reduce parameter dependencies in fitting procedures. Beyond temporal correlations, the characterization of spatial noise correlations presents an exciting direction for future research. Models of spatial correlation functions, such as those proposed for ferromagnetic materials \cite{Ziffer2024}, could be experimentally validated by combining spatial and temporal noise measurements. Additionally, while pure dephasing models provide a natural baseline for quantum sensing, an important extension would be to investigate scenarios where the fluctuating field is not aligned with the N$\!\rightarrow$V axis. In such cases, transverse components of the noise could induce relaxation and transitions of the NV center, expanding the applicability of the present framework to mixed dephasing–relaxation dynamics. This integrated approach would provide deeper insights into noise dynamics, paving the way for improved noise models and better control over quantum systems.
 
 Overall, this work provides a general and experimentally validated methodology to classify dynamical noise in quantum systems. It enables the detection of (oscillating) non-Markovian and/or nonstationary behavior without fully reconstructing the environmental correlation function. It addresses a key gap in current quantum sensing technology—namely, the lack of frameworks to disentangle and identify the interplay between non-Markovian and non-stationary features— by offering a practical and operationally motivated framework, and paves the way for robust, sensor-based characterization strategies in real-world quantum technologies.
\label{sec: conclusion}

\section*{Declarations}

\begin{itemize}

\item Data availability \\
The data relevant to figures in the main text are available via the Weizmann/Elsevier data repository, DOI: 10.34933/e08be44a-98e7-4d1e-9faf-d36205c1dd27~\cite{Rosenberg2025a}. Additional raw data are available from the corresponding author upon reasonable request.

\item Code availability \\
The python code is available from the authors upon reasonable request.

\begin{acknowledgments}
We acknowledge fruitful discussions with Oren Raz on this manuscript, and thank Yoav Romach for his helpful advice on designing the code that controls the Quantum Machines OPX (v1). AF is the incumbent of the Elaine Blond Career Development Chair in Perpetuity and acknowledges funding from the Minerva Stiftung (Grant No.\,151018), and also support by the Kimmel Institute for Magnetic Resonance. This research is made possible in part by the historic generosity of the Harold Perlman Family. GA and AZ acknowledge support by CNEA; Fundaci\'{o}n Balseiro; CONICET; PIBBA, PICT-2018-4333; PICT-2021-GRF-TI-00134; PICT-2021-I-A-00070; UNCUYO SIIP Tipo I 2022-C002, 2022-C030; Instituto Balseiro; Collaboration program between the MINCyT (Argentina) and and MOST (Israel); the Morris Belkin Visiting Professorship and the Benoziyo Endowment Fund for the Advancement of Science. 
\end{acknowledgments}

\end{itemize}

\appendix

\renewcommand{\theequation}{\thesection\arabic{equation}}
\setcounter{equation}{0}

\section{Derivation of correlation functions}
\label{sec:Derivation_correlation_functions}
\subsection{Markovian noise correlation functions}
\label{sec:Markovian_correlation_functions}

For the Markovian ($m=0$) case, $n'(t)$ depends only on the state of the system at the given time $t$, and the trajectories of the system must satisfy the equation $n'+t^{-1}_cn=\frac{1}{\Gamma}\eta (t)$, where $t_c=\frac{\Gamma}{\kappa }$ is the correlation time. The effect of the initial state is described by the homogeneous solutions $n_0(t)=n_0e^{-\frac{t}{t_c}}$, where the initial state $n_0$ is independent of the driving---$\left\langle n_0\eta (t)\right\rangle =0$.  To find the trajectory for a particular $\eta(t)$, we can use the Green's function $G\left(t\right)=\mathit{\Theta}(t)e^{-\frac{t}{t_c}}$, which has the property $G'+t_c^{-1}G=\delta (t)$. The trajectory is given by $n\left(t\right)=n_0(t)+\int^t_0{G\left(t-s\right)\frac{1}{\Gamma}\mathrm{\ }\eta \left(s\right)ds}$, where the homogeneous part $n_0(t)$ is uniquely determined by the initial condition $n\left(0\right)=n_0$.  Now recall that the correlation function is defined by the expectation value $\left\langle n\left(t_1\right)n\left(t_2\right)\right\rangle $ for any $t_1,t_2$ on the same trajectory. Thus, we can treat $n_0$ as constant when taking the expectation value over a trajectory and afterwards take an average over a distribution of initial conditions to obtain
\begin{widetext}
\begin{equation} 
\begin{aligned}
\left\langle n\left(t_1\right)n(t_2)\right\rangle  & =\left\langle \left(n_0(t_1)+\int^{t_1}_0{G\left(t_1-s_1\right)\frac{1}{\Gamma}\eta \left(s_1\right)ds_1}\right)\left(n_0(t_2)+\int^{t_2}_0{G\left(t_2-s_2\right)\frac{1}{\Gamma}\eta \left(s_2\right)ds_2}\right)\right\rangle  \\
& = \left\langle n_0\left(t_1\right)n_0(t_2)\right\rangle +\int^{t_1}_0{\int^{t_2}_0{G\left(t_1-s_1\right)G\left(t_2-s_2\right)\frac{1}{{\Gamma}^2}\left\langle \eta \left(s_1\right)\eta \left(s_2\right)\right\rangle ds_2}ds_1}
\end{aligned}
\end{equation}
\end{widetext}
Note that this correlation function has two parts: the homogeneous part $\left\langle n_0\left(t_1\right)n_0(t_2)\right\rangle $, which depends only on the initial condition $n_0$, and the heterogeneous part, which depends only on the correlation function of the driving force.  This general expression can be used for any system for which the Green's function is known.  Using the homogeneous solution and Green's function for a Markovian system driven by white noise, we get the homogeneous part $\left\langle n_0\left(t_1\right)n_0(t_2)\right\rangle =\left\langle n^2_0\right\rangle e^{-\frac{t_1+t_2}{t_c}}\mathrm{\ }$and the heterogeneous part
\begin{widetext}
\begin{equation}
\begin{aligned}
\left\langle n\left(t_1\right)n(t_2)\right\rangle -\left\langle n_0\left(t_1\right)n_0\left(t_2\right)\right\rangle & =\int^{t_1}_0{\int^{t_2}_0{e^{-\frac{t_1-s_1}{t_c}}e^{-\frac{t_2-s_2}{t_c}}\frac{A}{{\Gamma}^2}\delta \left(s_1-s_2\right)ds_2}ds_1} \\
& =\int^{{\mathrm{min} \left(t_1,t_2\right)\ }}_0{\frac{A}{{\Gamma}^2}e^{-\frac{t_1+t_2-2s}{t_c}}ds}=\frac{At_c}{2{\Gamma}^2}\left(e^{-\frac{\left|t_1-t_2\right|}{t_c}\ }-e^{-\frac{t_1+t_2}{t_c}}\right)
\end{aligned}
\end{equation}
\end{widetext}
so the correlation function for Markovian noise is
\begin{equation}
\left\langle n\left(t_1\right)n(t_2)\right\rangle =\left\langle n^2_0\right\rangle e^{-\frac{t_1+t_2}{t_c}}+{\mathrm{\Delta }}^2\left(e^{-\frac{\left|t_1-t_2\right|}{t_c}}-e^{-\frac{t_1+t_2}{t_c}}\right)\,,
\end{equation}

where ${\mathrm{\Delta }}^{\mathrm{2}}={\lim_{t\to \infty } \left\langle {n(t)}^2\right\rangle \ }=\frac{At_c}{2{\Gamma}^2}$ is the variance at equilibrium. At equilibrium, the correlation function is 
\begin{equation}
{\lim_{t_1,t_2\to \infty } \left\langle n\left(t_1\right)n(t_2)\right\rangle ={\mathrm{\Delta }}^2e^{-\frac{\left|t_1-t_2\right|}{t_c}}\ }\,.
\end{equation}
If we quench the system so that $n_0=0$ at the start of each measurement, then the correlation function is 
\begin{equation}{\left\langle n\left(t_1\right)n(t_2)\right\rangle }_{n_0=0}={\mathrm{\Delta }}^2\left(e^{-\frac{\left|t_1-t_2\right|}{t_c}}-e^{-\frac{t_1+t_2}{t_c}}\right)\,.
\end{equation}
Note that if we average over an ensemble with $\left\langle {n_0}^2\right\rangle ={\mathrm{\Delta }}^2$ , then the homogeneous part will exactly cancel the transient term. If we quench the system some time $t_d$ before the start of the measurement, then only the non-stationary term will be affected by the delay, so the correlation function then becomes
\begin{equation}
\left\langle n\left(t_1\right)n(t_2)\right\rangle ={\mathrm{\Delta }}^2\left(e^{-\frac{\left|t_1-t_2\right|}{t_c}}-e^{-\frac{t_1+t_2+2t_d}{t_c}}\right)\,.    
\end{equation}
The functional form of this correlation function leads us to the (somewhat surprising) conclusion that a Markovian noise bath is characterized by only two processes: fluctuations with correlation time $t_c$, and dissipations of (externally imposed) quenches with a decay time of $\frac{t_c}{2}$.  The fact that the Fourier transform of the equilibrium term of the Markovian noise correlation function is a Lorentzian explains the ubiquity of Lorentzian noise in nature. 
\subsection{{Underdamped non-Markovian noise correlation functions}} \label{sec:Underdamped_correlation_functions}

 One can use the same method to obtain correlation functions for non-Markovian noise.  In particular, we are interested in the $m\neq 0$ case, where the evolution of the noise and its derivative is Markovian, and thus the initial state is described by the initial values (at $t=0$) of the noise $n_0$ and that of its derivative $n_0'$, and the trajectories of the system satisfy the equation $n''+2t^{-1}_cn'+{\omega_0 }^2n=\frac{1}{m}\eta (t)$. Depending on the parameters, there are three possible functional forms for the homogeneous solutions $n_0(t)$.  When $\omega_0 t_c>1$, the system is underdamped, and the homogeneous solution is 

\begin{equation}
 n_0(t)=\left(n'_0+t^{-1}_cn_0\right)\frac{1}{\Omega}{\mathrm{sin} \left(\Omega t\right)\ }e^{-\frac{t}{t_c}}+n_0{\mathrm{cos} (\Omega t)\ }e^{-\frac{t}{t_c}}\,
 \label{eq:G_hom_non_Mark}
 \end{equation}
 where $\Omega=\sqrt{{\omega_0 }^2-t^{-2}_c}$. When $\omega_0 t_c<1$, the system is overdamped, and the homogeneous solution is 
 \begin{equation}
 n_0(t)=\left(n'_0+t^{-1}_cn_0\right)\frac{1}{\alpha }{\mathrm{sinh} \left(\alpha t\right)\ }e^{-\frac{t}{t_c}}+n_0{\mathrm{cos} (\alpha t)\ }e^{-\frac{t}{t_c}}\,,
 \end{equation}
 where $\alpha =\sqrt{t^{-2}_c-{\omega_0 }^2}$. When $\omega_0 t_c=1$, the system is critically damped, and the homogeneous solution is
 \begin{equation}
 n_0(t)=\left(n'_0+t^{-1}_cn_0\right)te^{-\frac{t}{t_c}}+n_0e^{-\frac{t}{t_c}}\,.
 \end{equation}
 For the experiments presented in this paper, we restrict ourselves to the underdamped case---which is a typical description of a nuclear spin bath.  For the underdamped case, we have the continuous Green's function $G\left(t\right)=\mathit{\Theta}(t)\frac{1}{\Omega}\mathrm{sin}\mathrm{}(\Omega t)e^{-\frac{t}{t_c}}$.  Note that, as before, the homogeneous part of the correlation function $\left\langle n_0\left(t_1\right)n_0(t_2)\right\rangle $ decays to zero for large $t_1,t_2$, so the equilibrium terms are all contained in the heterogeneous part. For the heterogeneous part, we have
 \begin{widetext}
 \begin{equation}
 \begin{aligned}
 & \left\langle n\left(t_1\right)n(t_2)\right\rangle -\left\langle n_0\left(t_1\right)n_0(t_2)\right\rangle \\
 & =\int^{t_1}_0\int^{t_2}_0{\frac{1}{\Omega^2}}{\mathrm{sin} \left(\Omega\left(t_1-s_1\right)\right)\ }e^{-\frac{t_1-s_1}{t_c}\ }{\mathrm{sin} \left(\Omega\left(t_2-s_2\right)\right)\ }e^{-\frac{t_2-s_2}{t_c}}\frac{A}{m^2}\delta \left(s_1-s_2\right)ds_2ds_1\\
 &=\frac{A}{2{\Omega}^2m^2}\int^{{\mathrm{min} \left(t_1,t_2\right)\ }}_0{e^{-\frac{t_1+t_2-2s}{t_c}}\left({\mathrm{cos} \left(\Omega\left(t_1-t_2\right)\right)\ }-{\mathrm{cos} \left(\Omega\left(t_1+t_2-2s\right)\right)\ }\right)ds}\\
 &={\mathrm{\Delta }}^2\left(\left({\mathrm{cos} \left(\Omega\left(t_1-t_2\right)\right)\ }+\frac{\mathrm{1}}{\mathrm{\Omega }t_c}{\mathrm{sin} \left(\Omega\left|t_1-t_2\right|\right)\ }\right)e^{-\frac{\left|t_1-t_2\right|}{t_c}}-\left(\frac{{\omega_0 }^2}{{\Omega}^2}{\mathrm{cos} \left(\Omega\left(t_1-t_2\right)\right)\ }\right.\right.\\
 &\left.\left.-\frac{1}{{\Omega}^2t^2_c}{\mathrm{cos} \left(\Omega\left(t_1+t_2\right)\right)\ }+\frac{\mathrm{1}}{\Omega t_c}{\mathrm{sin} \left(\Omega\left(t_1+t_2\right)\right)\ }\right)e^{-\frac{t_1+t_2}{t_c}}\right)
 \end{aligned}
 \label{eq:G_het_non_Mark}
 \end{equation}
where ${\mathrm{\Delta }}^2={\lim_{t\to \infty } \left\langle {n\left(t\right)}^2\right\rangle \ }=\frac{At_c}{4{\omega_0 }^2m^2}=\frac{At_c}{4m^2\left({\mathrm{\Omega }}^2+t^{-2}_c\right)}$ is the variance at equilibrium. Because the homogeneous part disappears in the limit of large $t_1,t_2$, the correlation function at equilibrium is
\begin{equation}
{\lim_{t_1,t_2\to \infty } \left\langle n\left(t_1\right)n\left(t_2\right)\right\rangle \ }={\mathrm{\Delta }}^2\left({\mathrm{cos} \left(\Omega\left(t_1-t_2\right)\right)\ }+\frac{1}{\Omega t_c}{\mathrm{sin} \left(\Omega\left|t_1-t_2\right|\right)\ }\right)e^{-\frac{\left|t_1-t_2\right|}{t_c}}\,.
\end{equation}
\end{widetext}
Also, if we quench the system so that $n_0=n'_0=0$, then the homogeneous part will vanish, and the correlation function will consist only of the heterogeneous part. To calculate the equilibrium variance of the derivative ${\lim_{t\to \infty } \left\langle {n'\left(t\right)}^2\right\rangle \ }$ and the equilibrium covariance ${\lim_{t\to \infty } \left\langle n'\left(t\right)n\left(t\right)\right\rangle \ }$, we can use the fact that the correlation function at equilibrium is twice differentiable everywhere (even where $t_1=t_2$). We then obtain 

\begin{equation}
\begin{aligned}
{\lim_{t\to \infty } \left\langle n'\left(t\right)n\left(t\right)\right\rangle \ }&=\left.\left[\frac{\partial }{\partial t_1}{\lim_{t_1,t_2\to \infty } \left\langle n\left(t_1\right)n\left(t_2\right)\right\rangle \ }\right]\right|_{t_1=t_2}\\
=&\left.{-\Delta^2 \frac{\omega_0^2}{\Omega}{\mathrm{sin} \left(\Omega\left(t_1-t_2\right)\right)\ }e^{\frac{-|t_1-t_2|}{t_c}}\ }\right|_{t_1=t_2}\\
&=0\,,
\end{aligned}
\end{equation}
and
\begin{widetext}
\begin{equation}
\begin{aligned}
{\lim_{t\to \infty } \left\langle {n'\left(t\right)}^2\right\rangle \ }& =\left.\left[\frac{{\partial }^2}{\partial t_2\partial t_1}{\lim_{t_1,t_2\to \infty } \left\langle n\left(t_1\right)n\left(t_2\right)\right\rangle \ } \right]\right|_{t_1=t_2}\\
& = {\left. \Delta^2\omega_0^2\left({\mathrm{cos} \left(\Omega\left(t_1-t_2\right)\right)\ }-\frac{1}{\Omega t_c
}{\mathrm{sin} \left(\Omega\left|t_1-t_2\right|\right)\ }\right)e^{-\frac{\left|t_1-t_2\right|}{t_c}}\ \right|_{t_1=t_2}}\\
&={\Delta }^2{\omega_0 }^2\,.
\end{aligned}
\end{equation}
As before, when we replace these parameters in the homogeneous part, averaging over some ensemble with $\left\langle {n_0}^2\right\rangle ={\mathrm{\Delta }}^2$, $\left\langle {n'_0}^2\right\rangle ={\mathrm{\Delta }}^2\left({\mathrm{\Omega }}^2+t^{-2}_c\right)={\Delta }^2{\omega_0 }^2$, and $\left\langle n_0n'_0\right\rangle =0$, then the homogeneous part becomes 
\begin{equation}
\left\langle n_0(t_1)n_0(t_2)\right\rangle = {\mathrm{\Delta }}^2\left(\frac{1}{\mathrm{\Omega }t_c}{\mathrm{sin} \left(\Omega\left(t_1+t_2\right)\right)\ }+\frac{{\omega_0 }^2}{{\mathrm{\Omega }}^2}{\mathrm{cos} \left(\Omega\left(t_1-t_2\right)\right)\ }-\frac{1}{{\mathrm{\Omega }}^2t^2_c}{\mathrm{cos} \left(\Omega\left(t_1+t_2\right)\right)\ }\right)e^{-\frac{t_1+t_2}{t_c}}\,,
\end{equation}
\end{widetext}
which exactly cancels the transient terms in the heterogeneous part, leaving only the equilibrium terms. 

The functional form of $\left\langle n\left(t_1\right)n(t_2)\right\rangle $ for the noise of an underdamped non-Markovian spin bath shows a much richer behavior than for the Markovian case.  One key feature is that the oscillatory behavior of the spin bath persists even at equilibrium.  The Fourier transform of the noise from this bath will therefore be centered around the nonzero frequencies $\pm\sqrt{\Omega^2-t_c^{-2}}$ rather than around zero (see Appendix \ref{sec:Spectral_densities}).  

\subsection{Summary of correlation functions} \label{sec:summary_correlation_functions}

The correlation functions for equilibrium and quenched Markovian ($m=0$) and underdamped non-Markovian ($m\neq 0$) noise are summarized in Table \ref{tab:udnm}:

\noindent 
\begin{table*}[!ht]
\begin{tabular}{|p{1.8in}|p{3.5in}|} \hline 
\textbf{Noise Type} & \textbf{Correlation Function }$\left\langle \boldsymbol{n}\left({\boldsymbol{t}}_{\boldsymbol{1}}\right)\boldsymbol{n}\left({\boldsymbol{t}}_{\boldsymbol{2}}\right)\right\rangle $\textbf{} \\ \hline 
Equilibrium Markovian & ${\mathrm{\Delta }}^2e^{-\frac{\left|t_1-t_2\right|}{t_c}}$ \\ \hline 
Quenched Markovian & ${\mathrm{\Delta }}^2\left(e^{-\frac{\left|t_1-t_2\right|}{t_c}}-e^{-\frac{t_1+t_2}{t_c}}\right)$ \\ \hline 
Equilibrium underdamped non-Markovian & ${\mathrm{\Delta }}^2\left({\mathrm{cos} \left(\Omega\left(t_1-t_2\right)\right)\ }+\frac{1}{t_c\mathrm{\Omega }}{\mathrm{sin} \left(\Omega\left|t_1-t_2\right|\right)\ }\right)e^{-\frac{\left|t_1-t_2\right|}{t_c}}$ \\ \hline 
Quenched underdamped non-Markovian & {\begin{align*}{\mathrm{\Delta }}^2&\left(\left({\mathrm{cos} \left(\Omega\left(t_1-t_2\right)\right)\ }+\frac{\mathrm{1}}{\mathrm{\Omega }t_c}{\mathrm{sin} \left(\Omega\left|t_1-t_2\right|\right)\ }\right)e^{-\frac{\left|t_1-t_2\right|}{t_c}}\right.\\
& -\left(\frac{{\omega_0 }^2}{{\Omega}^2}{\mathrm{cos} \left(\Omega\left(t_1-t_2\right)\right)\ }-\frac{1}{{\Omega}^2t^2_c}{\mathrm{cos} \left(\Omega\left(t_1+t_2\right)\right) }\right.\\
&\left.\left.+\frac{\mathrm{1}}{\Omega t_c}{\mathrm{sin} \left(\Omega\left(t_1+t_2\right)\right)\ }\right)e^{-\frac{t_1+t_2}{t_c}}\right)\end{align*}} \\ \hline 
\end{tabular}
\caption{Table of correlation functions for equilibrium and quenched ($n_0=0$) Markovian noise, as well as for equilibrium and quenched ($n_0=n'_0=0$) underdamped non-Markovian noise.}
\label{tab:udnm}
\end{table*}

\section{Derivation of Ramsey decay exponents} \label{sec:derivation_Ramsey_exponents}

\subsection{Markovian noise Ramsey decay exponents} \label{sec:Markovian_noise_Ramsey_exponents}

For equilibrium Markovian noise, the correlation function is $\left\langle n\left(t_1\right)n\left(t_2\right)\right\rangle ={\Delta }^2e^{\frac{-\left|t_1-t_2\right|}{t_c}}$.  Thus, the time dependence of the exponent is 
\begin{equation}
\begin{aligned}
{\chi }_{m=0,eq}\left(t\right)=&\int^t_0{\int^{t_1}_0{{\Delta }^2e^{\frac{-\left|t_1-t_2\right|}{t_c}}dt_2}dt_1}\\
=&{\Delta }^2t_ct-{\Delta }^2{t_c}^2\left(1-e^{\frac{-t}{t_c}}\right).
\label{eq:S15}
\end{aligned}
\end{equation}

For short times $t\ll t_c$, we can approximate ${\chi }_{m=0,eq}\left(t\right)\approx \frac{{\Delta }^2}{2}t^2-\frac{{\Delta }^2}{6t_c}t^3+O(t^4)$, and for long times $t\gg t_c$, we can approximate ${\chi }_{m=0,eq}\left(t\right)\approx {\Delta }^2t_ct-{\Delta }^2{t_c}^2+O(e^{\frac{-t}{t_c}})$.

For quenched Markovian noise, we can write the correlation function as 
\begin{equation}
\left\langle n\left(t_1\right)n\left(t_2\right)\right\rangle ={\Delta }^2\left(e^{\frac{-\left|t_1-t_2\right|}{t_c}}-e^{\frac{-t_1+t_2}{t_c}}\right)\,.
\end{equation}
Thus, the time dependence of the exponent is
\begin{equation}
\begin{aligned}
{\chi }_{m=0,qu}\left(t\right)&=\int^t_0{\int^{t_1}_0{{\Delta }^2\left(e^{\frac{-\left|t_1-t_2\right|}{t_c}}-e^{\frac{-t_1+t_2}{t_c}}\right)dt_2}dt_1}\\
&={\Delta }^2t_ct-{\Delta }^2{t_c}^2\left({{\frac{3}{2}}}-2e^{\frac{-t}{t_c}}+{{\frac{1}{2}}}e^{\frac{-2t}{t_c}}\right)
\label{eq:S17}
\end{aligned}
\end{equation}
For short times $t\ll t_c$, we can approximate ${\chi }_{m=0,qu}\left(t\right)\approx \frac{{\Delta }^2}{3t_c}t^3+O\left(t^4\right)=\frac{A}{6{\Gamma}^2}t^3+O\left(t^4\right)\ $, and for long times $t\gg t_c$, we can approximate ${\chi }_{m=0,qu}\left(t\right)\approx {\Delta }^2t_ct-\frac{3}{2}{\Delta }^2{t_c}^2+O(e^{\frac{-t}{t_c}})$. Note how the leading order of the short time exponent for the quenched Markovian noise is proportional to the (normalized) driving strength $\frac{A}{{\Gamma}^2}$.

For Markovian noise with a quench delay, we can write the correlation function as 

\begin{equation}\left\langle n\left(t_1\right)n\left(t_2\right)\right\rangle ={\Delta }^2\left(e^{\frac{-\left|t_1-t_2\right|}{t_c}}-Qe^{\frac{-t_1+t_2}{t_c}}\right),
\end{equation}
with $\,\, Q=e^{\frac{-2t_d}{t_c}}$. Thus, the linearity of the double integral gives us the time dependence of the exponential as
\begin{equation}
{\chi }_{m=0,del}\left(t;t_d\right)=\left(1-Q\right){\chi }_{m=0,eq}\left(t\right)+Q{\chi }_{m=0,qu}\left(t\right)\,.
\label{eq:S19}
\end{equation}
This indicates that one can use the Ramsey signal to determine the time of a quench relative to the start of the Ramsey experiment.

For Markovian noise with a switch in correlation time, where the correlation time abruptly switches from $t_a$ to $t_b$ at some time $t_s$, we can write the correlation function as 

\begin{equation}
\left\langle n\left(t_1\right)n\left(t_2\right)\right\rangle =\left\{ \begin{array}{c}
 \begin{array}{c}
{\Delta }^2e^{\frac{-\left|t_1-t_2\right|}{t_a}}\ \text{for}\ t_1,t_2\le t_s \\ 
{\Delta }^2e^{\frac{-t_s-t_1}{t_a}-\frac{t_2-t_s}{t_b}}\ \text{for}\ t_1<t_s<t_2 \end{array}
 \\ 
 \begin{array}{c}
{\Delta }^2e^{\frac{-t_s-t_2}{t_a}-\frac{t_1-t_s}{t_b}}\ \text{for}\ t_2<t_s<t_1 \\ 
{\Delta }^2e^{\frac{-\left|t_1-t_2\right|}{t_b}}\ \text{for}\ t_1,t_2\ge t_s \end{array}
 \end{array}
\right.
\end{equation}
Thus, the time dependence of the exponent will also change abruptly at $t_s$.  For $t\le t_s$, we have

\begin{equation}
\begin{aligned}
{\chi }_{m=0,sw,\ t\le t_s}\left(t\right)&=\int^t_0{\int^{t_1}_0{{\Delta }^2e^{\frac{-\left|t_1-t_2\right|}{t_a}}dt_2}dt_1}\\
&={\Delta }^2t_at-{\Delta }^2{t_a}^2\left(1-e^{\frac{-t}{t_a}}\right)
\label{eq:S21}
\end{aligned}
\end{equation}
\begin{widetext}
For $t>t_s$, we have to split the double integral into three regions, and we have
\begin{equation}
\begin{aligned}
{\chi }& _{m=0,sw,\ t>t_s}\left(t\right)\\
& =\int^{t_s}_0{\int^{t_1}_0{{\Delta }^2e^{\frac{-\left|t_1-t_2\right|}{t_a}}dt_2}dt_1}+\int^t_{t_s}{\int^{t_1}_{t_s}{{\Delta }^2e^{\frac{-\left|t_1-t_2\right|}{t_a}}dt_2}dt_1}+\int^t_{t_s}{\int^{t_s}_0{{\Delta }^2e^{\frac{-t_s-t_2}{t_a}-\frac{t_1-t_s}{t_b}}dt_2}dt_1}\\
& ={\Delta }^2t_at_s-{\Delta }^2{t_a}^2\left(1-e^{\frac{-t_s}{t_a}}\right)+{\Delta }^2t_b\left(t-t_s\right)-{\Delta }^2{t_b}^2\left(1-e^{\frac{-t-t_s}{t_b}}\right)\\
&+{\Delta }^2t_at_b\left(1-e^{\frac{-t_s}{t_a}}\right)\left(1-e^{\frac{-t-t_s}{t_b}}\right)
\end{aligned}
\label{eq:S22}
\end{equation}
This indicates that one can use the Ramsey signal to determine the time of an abrupt change in the correlation time, provided one acquires enough time points before and after the switch.

\noindent 

\subsection{Underdamped non-Markovian Ramsey decay exponents} \label{sec:underdamped_Ramsey_exponents}

For equilibrium underdamped non-Markovian ($m\neq 0$) noise, the correlation function is 
\begin{equation}
\left\langle n\left(t_1\right)n\left(t_2\right)\right\rangle ={\mathrm{\Delta }}^{\mathrm{2}}\left({\mathrm{cos} \left(\Omega\left(t_1-t_2\right)\right)\ }+\frac{1}{\mathrm{\Omega }t_c}\mathrm{sin}\mathrm{}\left(\Omega\left|t_1-t_2\right|\right)\right)e^{-\frac{\left|t_1-t_2\right|}{t_c}}\,,
\end{equation}
where $\Omega=\sqrt{{\omega_0 }^2-t^{-2}_c}$ is the damped frequency of the oscillator. Thus, the time dependence of the exponent is 
\begin{equation}
\begin{aligned}
{\chi }_{m\neq 0,eq}\left(t\right)&=\int^t_0{\int^{t_1}_0{{\mathrm{\Delta }}^2\left({\mathrm{cos} \left(\Omega\left(t_1-t_2\right)\right)\ }+\frac{1}{\mathrm{\Omega }t_c}\mathrm{sin}\mathrm{}\left(\Omega\left|t_1-t_2\right|\right)\right)e^{-\frac{\left|t_1-t_2\right|}{t_c}}dt_2}dt_1}\\
&=\frac{2{\mathrm{\Delta }}^2}{t_c{\omega_0 }^2}\ t-\frac{{\mathrm{\Delta }}^2}{{\omega_0 }^4}\left(\left({\Omega}^2-3t^{-2}_c\right){\mathrm{cos} \left(\Omega t\right)\ }+\frac{\left({3\Omega}^2-t^{-2}_c\right)}{\Omega t_c}{\mathrm{sin} \left(\Omega t\right)\ }\right)e^{-\frac{t}{t_c}}+\frac{{\Delta }^2\left({\Omega}^2-3t^{-2}_c\right)}{{\omega_0 }^4}
\label{eq:S24}
\end{aligned}
\end{equation}
For short times $t\ll t_c,\frac{1}{\Omega}$, we can approximate ${\chi }_{m\neq 0,eq}\left(t\right)\approx \frac{{\mathrm{\Delta }}^2}{2}t^2-\frac{{\Delta }^2{\omega_0 }^2}{24}t^4+\frac{{\Delta }^2{\omega_0 }^2}{60t_c}t^5+O(t^6)$, and for long times $t\gg t_c,\frac{1}{\Omega}$, we can approximate ${\chi }_{m\neq 0,eq}\left(t\right)\approx \frac{2{\Delta }^2}{t_c{\omega_0 }^2}t+\frac{{\Delta }^2\left({\Omega}^2-3t^{-2}_c\right)}{{\omega_0 }^4}+O\left(e^{-\frac{t}{t_c}}\right)$.

For quenched underdamped non-Markovian noise (when $n_0=n'_0=0$), the correlation function differs from the equilibrium correlation function by a transient term
\begin{equation}
\begin{aligned}
{\left\langle n\left(t_1\right)n\left(t_2\right)\right\rangle }_{tr}&=\left\langle n\left(t_1\right)n\left(t_2\right)\right\rangle -{\left\langle n\left(t_1\right)n\left(t_2\right)\right\rangle }_{eq}\\
&=-{\mathrm{\Delta }}^2\left(\frac{{\omega_0 }^2}{{\Omega}^2}{\mathrm{cos} \left(\Omega\left(t_1-t_2\right)\right)\ }-\frac{1}{{\Omega}^2t^2_c}{\mathrm{cos} \left(\Omega\left(t_1+t_2\right)\right)\ }+\frac{\mathrm{1}}{\Omega t_c}{\mathrm{sin} \left(\Omega\left(t_1+t_2\right)\right)\ }\right)e^{-\frac{t_1+t_2}{t_c}}
\end{aligned}
\end{equation}
This transient term changes the time dependence of the exponent by
\begin{equation}
\begin{aligned}
{\chi }_{m\neq 0,tr}\left(t\right) =&-{\mathrm{\Delta }}^2\int^t_0\int^{t_1}_0e^{-\frac{t_1+t_2}{t_c}}\\
&\left(\frac{{\omega_0 }^2}{{\Omega}^2}{\mathrm{cos} \left(\Omega\left(t_1-t_2\right)\right)\ }-\frac{1}{{\Omega}^2t^2_c}{\mathrm{cos} \left(\Omega\left(t_1+t_2\right)\right)\ }+\frac{\mathrm{1}}{\Omega t_c}{\mathrm{sin} \left(\Omega\left(t_1+t_2\right)\right)\ }\right)dt_2dt_1 \\
=&-\frac{{\mathrm{\Delta }}^2}{{\omega_0 }^4}\left[\left(\frac{{\omega_0 }^4}{2{\mathrm{\Omega }}^2}+\frac{\left(3{\Omega}^2-t^{-2}_c\right)}{2{{\mathrm{\Omega }}^2t}^2_c}{\mathrm{cos} \left(2\Omega t\right)\ }+\frac{\left(3t^{-2}_c-{\Omega}^2\right)}{2\mathrm{\Omega }t_c}{\mathrm{sin} \left(2\Omega t\right)\ }\right)e^{-\frac{2t}{t_c}}\right.\\
& +\left(-\left({\Omega}^2+5t^{-2}_c\right){\mathrm{cos} \left(\Omega t\right)\ }+\frac{\left({\Omega}^2-3t^{-2}_c\right)}{\mathrm{\Omega }t_c}{\mathrm{sin} \left(\Omega t\right)\ }\right)e^{-\frac{t}{t_c}} \\
& \left.-\left(\frac{{\omega_0 }^4}{2{\mathrm{\Omega }}^2}+\frac{\left(3{\Omega}^2-t^{-2}_c\right)}{2{{\mathrm{\Omega }}^2t}^2_c}-\left({\Omega}^2+5t^{-2}_c\right)\right)\right]
\end{aligned}
\label{eq:S26}
\end{equation}
\end{widetext}
For short times $t\ll t_c,\Omega^{-1}$, we can approximate ${\chi }_{N=2,tr}\left(t\right)\approx -\frac{{\mathrm{\Delta }}^2}{2}t^2+\frac{{\Delta }^2{\omega_0 }^2}{24}t^4+\frac{{\Delta }^2{\omega_0 }^2}{12t_c}t^5+O(t^6)$, so we have ${\chi }_{m\neq 0,qu}\left(t\right)={\chi }_{m\neq 0,eq}\left(t\right)+{\chi }_{m\neq 0,tr}\left(t\right)\approx \frac{{\Delta }^2{\omega_0 }^2}{10t_c}t^5+O\left(t^6\right)=\frac{A}{40m^2}t^5+O\left(t^6\right)$, and for long times $t\gg t_c,\Omega^{-1}$, we can approximate ${\chi }_{m\neq 0,tr}\left(t\right)\approx \frac{{\mathrm{\Delta }}^2}{{\omega_0 }^4}\left(\frac{{\omega_0 }^4}{2{\mathrm{\Omega }}^2}+\frac{\left(3{\Omega}^2-t^{-2}_c\right)}{2{{\mathrm{\Omega }}^2t}^2_c}-\left({\Omega}^2+5t^{-2}_c\right)\right)+O\left(e^{-\frac{t}{t_c}}\right)$, so we have ${\chi }_{m\neq 0,qu}\left(t\right)={\chi }_{m\neq 0,eq}\left(t\right)+{\chi }_{m\neq 0,tr}\left(t\right)\approx \frac{2{\Delta }^2}{t_c{\omega_0 }^2}t+\frac{{\mathrm{\Delta }}^2}{2{\Omega}^2}+\frac{{\mathrm{\Delta }}^2\left(3{\Omega}^2-t_c^{-2}\right)}{2{\Omega}^2t^2_c{\omega_0 }^4}\ -\frac{8{\mathrm{\Delta }}^2}{t^2_c{\omega_0 }^4}+O\left(e^{-\frac{t}{t_c}}\right)$. Note how the leading order of the short time exponent for the quenched non-Markovian noise is proportional to the (normalized) driving strength $\frac{A}{{m}^2}$.

\noindent 

\subsection{Summary of Ramsey decay exponents}
\label{sec:summary_Ramsey_exponents}
The short-time and long-time approximations of the Ramsey decay exponents for equilibrium and quenched Markovian ($m=0$) and underdamped non-Markovian ($m\neq 0$) noise are summarized in the Table \ref{tab:S2}:

\begin{table*}[!ht]
\begin{tabular}{|p{2.0in}|p{2.0in}|p{2.0in}|} \hline 
\textbf{Noise type} & \textbf{Ramsey exponent }$\chi \boldsymbol{(}\boldsymbol{t}\boldsymbol{)}$\textbf{, short }$\boldsymbol{t}$\textbf{} & \textbf{Ramsey exponent }$\chi \boldsymbol{(}\boldsymbol{t}\boldsymbol{)}$\textbf{, long }$\boldsymbol{t}$ \\ \hline 
Equilibrium Markovian\newline  &  $\chi \left(t\right)\approx \frac{{\Delta }^2}{2}t^2-\frac{{\Delta }^2}{6t_c}t^3+O(t^4)$ &  $\chi (t)\approx {\Delta }^2t_ct-{\Delta }^2{t_c}^2+O(e^{\frac{-t}{t_c}})$ \\ \hline 
Quenched Markovian\newline  &  $\chi \left(t\right)\approx \frac{{\Delta }^2}{3t_c}t^3+O(t^4)\ =\frac{1}{6}\frac{A}{{\Gamma}^2}t^3+O\left(t^4\right)$ &  $\chi (t)\approx {\Delta }^2t_ct-\frac{3}{2}{\Delta }^2{t_c}^2+O(e^{\frac{-t}{t_c}})$ \\ \hline 
Equilibrium underdamped non-Markovian &  $\chi \left(t\right)\approx \frac{{\mathrm{\Delta }}^2}{2}t^2-\frac{{\Delta }^2{\omega_0 }^2}{24}t^4+\frac{{\Delta }^2{\omega_0 }^2}{60t_c}t^5+O\left(t^6\right)$ &  $\chi (t)\approx \frac{{\mathrm{\Delta }}^2}{t_c{\left({\mathrm{\Omega }}^2+t^{-2}_c\right)}^2}t+{\mathrm{\Delta }}^2\frac{{\Omega}^2-3t^{-2}_c}{{\left({\mathrm{\Omega }}^2+t^{-2}_c\right)}^2}+O\left(e^{-\frac{t}{t_c}}\right)$ \\ \hline 
Quenched underdamped non-Markovian &  $\chi \left(t\right)\approx {\frac{{\Delta }^2{\omega_0 }^2}{10t_c}t}^5+O\left(t^6\right)=\frac{A}{40m^2}t^5+O\left(t^6\right)\mathrm{,\ }$ &  $\chi (t)\approx \frac{2{\Delta }^2}{t_c{\omega_0 }^2}t+\frac{{\mathrm{\Delta }}^2}{2{\Omega}^2}+\frac{{\mathrm{\Delta }}^2\left(3{\Omega}^2-t_c^{-2}\right)}{2{\Omega}^2t^2_c{\omega_0 }^4}\ -\frac{8{\mathrm{\Delta }}^2}{t^2_c{\omega_0 }^4}+O\left(e^{-\frac{t}{t_c}}\right)$ \\ \hline 
\end{tabular}
\caption{Table of short-time and long-time approximations of the Ramsey decay exponents $\chi(t)$ for the noise types described in Table\,\ref{tab:udnm}. The Ramsey decay is given by $\left\langle S_z\right\rangle =e^{-\chi (t)}$.}
\label{tab:S2}
\end{table*}

\subsection{Analytical plots of Ramsey decay curves for non-Markovian noise } \label{sec:Analytical_plots_nonMarkovian_noise}

\begin{figure}[!ht]
    \centering
    \includegraphics[width=\columnwidth]{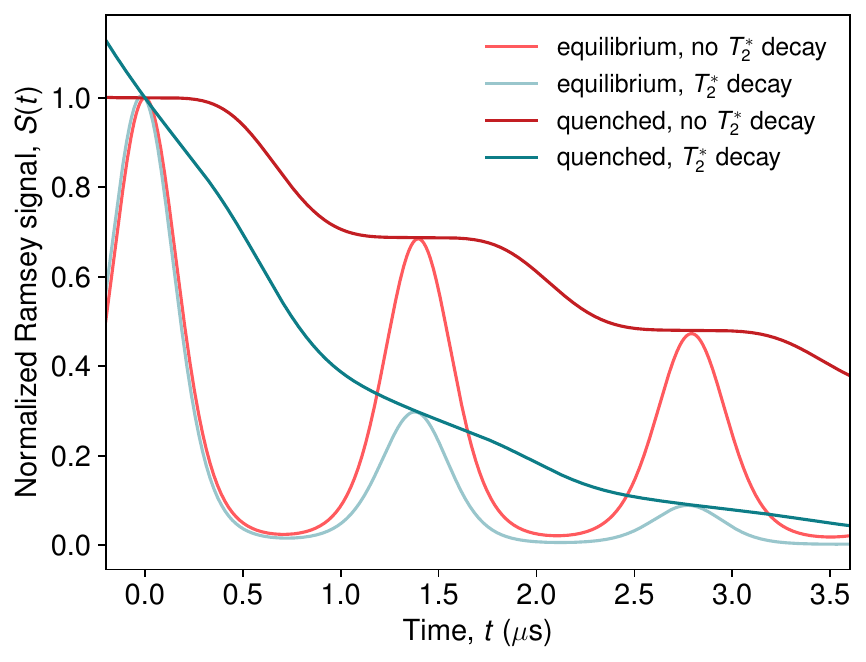}
    \caption{Analytical calculation of the Ramsey signal for equilibrium (light red, light blue) and quenched (red, blue) underdamped non-Markovian ($m\neq 0$) injected noise with a driving strength $A/m^2$ of 400\,MHz$^5$, restoring frequency 5\,MHz, and damping coefficient 0.2\,MHz, both without (red) and with (blue) an independent $T_2^*$ of 1.67\,$\upmu$s from non-injected white noise.}
    \label{fig: analytical}
\end{figure}

\noindent Plots of predicted Ramsey decay curves for quenched and equilibrium underdamped non-Markovian noise are shown in Fig.\,\ref{fig: analytical}. Note that, as discussed in the main text, the peak of the revival in the Ramsey decay curve for equilibrium underdamped non-Markovian noise always coincides with a minimum in the derivative of the Ramsey decay curve for quenched underdamped non-Markovian noise. When the quantum sensor is subject only to quenched underdamped non-Markovian noise, these plateaus at $t=\frac{2\pi m}{\Omega}$ for $m\in {\mathbb{Z}}^*$ are readily apparent. However, in our experiment, in addition to injected noise, there is also native noise from the diamond itself (see Fig.\,\ref{fig: ramsey_control}), which gives rise to a finite $T^*_2$ decay time even in the absence of injected noise. When the effect of this $T^*_2$ is included, the inflection points in the Ramsey decay curve for quenched underdamped non-Markovian noise become less apparent, as one can see by comparing the red and blue curves in Fig.\,\ref{fig: analytical}.

\noindent 

\subsection{A note on the parameters of $\chi(t)$ shown above}

Note that for the Markovian ($m=0$) case, we keep the equilibrium variance $\Delta_{m=0}^2$ constant and vary the correlation time $t_c$, so the equilibrium Ramsey decay remains constant and the quenched Ramsey decay becomes slower as $t_c$ increases (see Fig.\,\ref{fig: ramsey_eq_non_eq} in the main text), as predicted by the expressions for $\chi \left(t\right)$ shown above. For the non-Markovian ($m\neq 0$) case, we keep the driving strength $A/m^2$ constant and vary the self-correlation time $t_c$. Therefore, as $t_c$ increases, the equilibrium variance for the non-Markovian case $\Delta_{m\neq 0}^2=\frac{At_c}{4{\omega_0 }^2m^2}$ increases, so the equilibrium Ramsey decay becomes faster and the quenched Ramsey decay remains constant (see Fig.\,\ref{fig: nm} in the main text), as predicted by the expressions for $\chi \left(t\right)$ shown above, in Appendix \ref{sec:underdamped_Ramsey_exponents}.

\section{Equilibrium noise spectral densities for Markovian and underdamped noise}\label{sec:Spectral_densities}

Using the fact that the correlation functions for equilibrium noise are invariant to translations in time, we can define the equilibrium noise spectral density,
\begin{equation}
S(\omega) = \frac{1}{2\pi}\int_{-\infty}^\infty \langle n(t)n(0)\rangle e^{i\omega t}dt.
\end{equation}
For equilibrium Markovian noise, the noise spectral density is 
\begin{equation}
S_{m=0}(\omega)=\frac{\Delta^2}{2\pi}\int_{-\infty}^\infty e^{-|t|/t_c + i\omega t} dt = \frac{\Delta^2}{\pi}\frac{t_c}{1+t_c^2\omega^2}
\end{equation}
which is peaked at $\omega=0$ and decays as $O(\omega^2)$ for large $\omega$. The fact that the noise spectrum of a Markovian noise process is Lorentzian explains the ubiquity of Lorentzian noise spectra in nature.
For equilibrium underdamped non-Markovian noise, the noise spectral density is
\begin{widetext}
\begin{equation}
S_{m\neq0}(\omega) = \frac{\Delta^2}{2\pi}\int_{-\infty}^{\infty}\left(\cos(\Omega t)+\frac{1}{t_c\Omega}\sin(|\Omega t|)\right)e^{-|t|/t_c + i\omega t}dt = \frac{2\Delta^2\omega_0^2}{\pi t_c}\frac{ 1}{\left(\omega^2-\omega_0^2\right)^2 + 4t_c^{-2}\omega^2}
\end{equation}
\end{widetext}
which is peaked at $\omega = \pm \sqrt{\omega_0^2 - 2t_c^{-2}}$ (or at $0$ if $\omega_0 t_c <\sqrt2$) and decays as $O(\omega^4)$ for large $\omega$. The lower spectral density for large $\omega$ results from the fact that the first derivative of a non-Markovian noise trajectory must be continuous.  Indeed, simply by comparing the trajectories in Fig.\,\ref{fig: setup}(b) and \ref{fig: setup}(c) in the main text, one can observe that the discontinuities in the first derivative of the Markovian noise trajectories give rise to stronger higher frequency components.

\section{Physical realizations of the $m\neq0$ model} \label{sec:physical_realizations}

In this Appendix we show physical realizations of the $m\neq0$ model. This shows how different physical cases, including quantum systems, can give rise to noise closely resembling that of classical stochastic processes.
\subsection{Damped quantum harmonic oscillator} \label{subsec:Damped_oscillator}
A possible application of the $m\neq0$ model is a quantum sensor coupled to the position of a quantum harmonic oscillator dampened due to interacting with a finite-temperature Caldeira-Legget bath with Ohmic spectral density. The evolution of the position $n(t)$ of this system is described by \cite{Breuer2002}
\begin{equation}
    n''(t) + 2t_c^{-1} n'(t) + \omega_0^2 n(t) = \frac{F(t)}{m}\,,
    \label{eq:Langevin-like-QHO}
\end{equation}
where $F(t)$ is the fluctuating force felt by the oscillator due to its environment. $F(t)$ need not be white. However, under certain conditions, such as high temperature, white noise can be recovered from this system. Moreover, when the harmonic oscillator is underdamped (i.e. weakly coupled to the Caldeira-Legget bath), its dynamics will act as a pass-band filter where the system can only exchange energy with the modes of frequency $\omega_0$. In this case, the correlation function of $F(t)$ becomes irrelevant, as only the value of its noise spectrum at $\omega_0$ will affect the oscillator dynamics. To see an example of this, consider the stationary noise spectrum of the oscillator \cite{Breuer2002}
\begin{equation}
    S(\omega) = \frac{2}{mt_c} \frac{\omega \, \coth\left( \frac{\omega}{2 T} \right)}{\left(\omega^2-\omega_0^2\right)^2 + 4t_c^{-2}\omega^2}\,,
\end{equation}
where $T$ is the temperature and $\hbar=k_B=1$.
Note that this noise spectrum does not match that of a classical $m\neq0$ process, as the term $\omega \, \coth\left( \frac{\omega}{2 T} \right)$ induces an extra dependence on the frequency. However, at high temperature, we can take $\omega \, \coth\left( \frac{\omega}{2 T} \right)\sim2T$ and obtain 
\begin{equation}
        S(\omega) = \frac{4 T}{t_cm} \frac{1}{\left(\omega^2-\omega_0^2\right)^2 + 4t_c^{-2}\omega^2}\,,
\end{equation}
which does match our model. But it is not necessary to take the high temperature limit, as if $\omega t_c\gg1$, the noise spectrum will be comprised of two peaks at $\omega=\pm \sqrt{\omega_0^2 - 2t_c^{-2}}$, in the same positions as the peaks of $S_{m\neq0}(\omega)$ in Appendix \ref{sec:Spectral_densities}. Thus, when studying the peaks of the spectrum we can approximate
\begin{equation}
        S(\omega) = \frac{2 }{t_cm} \omega_0 \, \coth\left( \frac{\omega_0}{2 T} \right) \frac{1}{\left(\omega^2-\omega_0^2\right)^2 + 4t_c^{-2}\omega^2}\,,
\end{equation}
which again matches the results of our model. As seen in Eq.\,(\ref{eq:Langevin-like-QHO}), the parameter $m$ of our model coincides with the physical mass of the harmonic oscillator in this case.

\subsection{Rotating spin bath}\label{subsec:simple_model}
We consider the case where the NV couples to a bath of environmental spins that, in addition to experiencing thermal fluctuations, also precess with frequency $\Omega$. To model this class of systems in the simplest possible terms, we consider that the environmental operator that induces dephasing in the NV is $n\propto I_x$, where
\begin{equation}
    \mathbf{I}=
    \begin{pmatrix}
    I_x\\I_y    
    \end{pmatrix}
\end{equation}
is the environmental magnetization, restricted to the direction perpendicular to Larmor precession. When the NV interacts with many environmental spins, the environmental field $\mathbf{I}$ may be replaced by a Gaussian, semi-classical random process. We describe this process with the stochastic differential equation
\begin{equation}
        \mathbf{I}'(t)=\begin{pmatrix}
          -t_{c}^{-1} & \Omega\\
          -\Omega  & -t_{c}^{-1}
          \end{pmatrix}\mathbf{I}(t) + 
          \boldsymbol{\eta}\,,
          \label{eq:SDE_simple_model}
\end{equation}
where $\Omega$ is the effective rotation frequency of the bath spins, $t_{c}^{-1}$ is the lifetime of their magnetization and $\boldsymbol{\eta}$ is a $2$D meanless Gaussian white noise process. Its correlation function is $\left\langle\boldsymbol{\eta}(t_1)\otimes\boldsymbol{\eta}(t_2)\right\rangle=A\delta(t_2-t_1)\mathbb{I}_2$, where $\mathbb{I}_2$ is the $2\times2$ identity matrix. 

To solve this equation, as before, use again Green functions, and write
\begin{equation}
    \mathbf{I}(t)=\mathbf{I}_0(t)+\int_0^t\mathbb{G}(t-s)\boldsymbol{\eta}(s)ds\,,
\end{equation}
where 
\begin{equation}
    \mathbf{I}_0(t)=e^{-\frac{t}{t_c}}
    \begin{pmatrix}
    \cos\left(\Omega t\right) &\sin\left(\Omega t\right) \\
    -\sin\left(\Omega t\right) &\cos\left(\Omega t\right)
    \end{pmatrix} \mathbf{I}_0
\end{equation}
is the homogeneous solution describing the decay of initial conditions and $\mathbb{G}$ is the Green function, that satisfies 
\begin{equation}
    \mathbb{G}'(t)-\begin{pmatrix}
          -t_{c}^{-1} & \Omega\\
          -\Omega  & -t_{c}^{-1}
          \end{pmatrix}\mathbb{G}(t)=\delta(t)\mathbb{I}_2\,.
\end{equation}
It is given by
\begin{equation}
    \mathbb{G}(t)=\Theta(t)e^{-\frac{t}{t_c}}
    \begin{pmatrix}
    \cos\left(\Omega t\right) &\sin\left(\Omega t\right) \\
    -\sin\left(\Omega t\right) &\cos\left(\Omega t\right)
    \end{pmatrix}\,.
\end{equation}

After performing the same procedure as in Appendix \ref{sec:Derivation_correlation_functions}, we obtain the correlation function for the noise process $n(t)$ to be
\begin{widetext}
\begin{equation}
\begin{aligned}
    \left\langle n(t_1) n(t_2)\right\rangle=&\Delta^2\mathrm{e}^{-\frac{\left|t_2-t_1\right|}{t_c}}\cos\left(\Omega\left(t_2-t_1\right)\right)
    \\+\mathrm{e}^{-\frac{t_1+t_2}{t_c}}&\left[ \left(\frac{\left\langle\mathbf{I}_0^2\right\rangle}{2}-\Delta^2\right) \cos\left(\Omega\left(t_1-t_2\right)\right) + \frac{\left\langle I_{0x}^2\right\rangle - \left\langle I_{0y}^2\right\rangle}{2} \cos\left(\Omega\left(t_1+t_2\right)\right)+\left\langle I_{0x}I_{0y}\right\rangle \sin\left(\Omega\left(t_1+t_2\right)\right) \right]\,,
    \label{eq:G_simple_model}
\end{aligned}
\end{equation}
where $\Delta^2$ is the equilibrium variance of $n(t)$. The homogeneous and heterogeneous parts are, respectively
\begin{subequations}
\label{eq:G_simple_spin_model}
\begin{equation}
\begin{aligned}
    \left\langle n_0(t_1) n_0(t_2)\right\rangle=&\mathrm{e}^{-\frac{t_1+t_2}{t_c}}\\
    \times&\left[ \frac{\left\langle\mathbf{I}_0^2\right\rangle}{2} \cos\left(\Omega\left(t_1-t_2\right)\right) + \frac{\left\langle I_{0x}^2\right\rangle - \left\langle I_{0y}^2\right\rangle}{2} \cos\left(\Omega\left(t_1+t_2\right)\right)+\left\langle I_{0x}I_{0y}\right\rangle\sin\left(\Omega\left(t_1+t_2\right)\right) \right]\,,
\end{aligned}
\end{equation}

\begin{equation}
    \left\langle n(t_1) n(t_2)\right\rangle - \left\langle n_0(t_1) n_0(t_2)\right\rangle =\Delta^2\left(\mathrm{e}^{-\frac{\left|t_2-t_1\right|}{t_c}}- \mathrm{e}^{\frac{t_1+t_2}{t_c}}\right)\cos\left(\Omega\left(t_2-t_1\right)\right)\,.
\end{equation}
\end{subequations}
\end{widetext}
Note the resemblance to Eqs.\,(\ref{eq:G_hom_non_Mark}) and (\ref{eq:G_het_non_Mark}). In particular, in the very underdamped limit, $\Omega t_c\to\infty$, the two expressions match, showing that our $m\neq0$ model recovers the behavior of this spin-bath model in the limit where the spins precess many times before decaying. However, since the correlation function is not smooth (as it has a cusp at $t_1=t_2$), the noise spectrum will be $O(\omega^2)$ and not $O(\omega^4)$. However, the oscillatory behavior in the self-correlation function will induce the same type of revivals in the Ramsey experiments as in the $m\neq0$ process. 
\subsection{Asymmetric spin bath}
\label{subsec:Asymmetric_spin_bath}
We also consider a more general process for modeling the environmental spins, where we replace Eq.\,(\ref{eq:SDE_simple_model}) by
\begin{equation}
    \mathbf{I}'(t)=\begin{pmatrix}
          -t_{cx}^{-1} & \omega_0\\
          -\omega_0  & -t_{cy}^{-1}
          \end{pmatrix}\mathbf{I}(t) + 
          \Sigma\boldsymbol{\eta}\,,
          \label{eq:SDE_complex_model}
\end{equation}
where $t_{cx}$ and $t_{cy}$ are the decay times in the $x$ and $y$ axes of the spin $I$, $\omega_0$ is the bare Larmor frequency and 
\begin{equation}
    \Sigma=\begin{pmatrix}
              \sigma_{x}&\sigma_{xy}\\
              \sigma_{xy}&\sigma_y        
    \end{pmatrix}
\end{equation}
is a $2\times2$ constant matrix that couples the white noise process $\boldsymbol{\eta}$ to the semi-classical process $\mathbf{I}$. This model includes anisotropy in the magnetization decay, as well as in the fluctuations. We focus in the limit where $t_{cx}^{-1},\,\sigma_x,\, \sigma_{xy}\to0$, and take $t_{cy}=\frac{t_c}{2}$, $\sigma_y=\left(m\omega_0\right)^{-1}$, $\eta=\boldsymbol{\eta}_y$. In this case, Eq.\,(\ref{eq:SDE_complex_model}) simplifies to
\begin{equation}
    \mathbf{I}'(t)=\begin{pmatrix}
          0 & \omega_0\\
          -\omega_0  & -2t_c^{-1}
          \end{pmatrix}\mathbf{I}(t) + 
          \begin{pmatrix}
          0\\\frac{\eta(t)}{m\omega_0}
          \end{pmatrix}\,.
\end{equation}
If we note that the first equation is just $I'_x(t)=\omega_0 I_y(t)$, after taking $n(t)= I_x(t)$, $n'(t)=\omega_0 I_y(t)$ we obtain
\begin{equation}
    n''(t)+2t_c^{-1}n'+\omega_0^2 n(t)=\eta(t)/m\,,
\end{equation}
which coincides with Eq.\,(\ref{eq: Langevin})
of the main text. Since same the stochastic differential equation describes both processes, the correlation functions of the processes are the same. In this case, the parameter $m=(\omega_0\sigma_y)^{-1}$ does not represent a physical mass but an effective quantity encapsulating the Larmor frequency and the coupling strength to other environmental degrees of freedom, modeled as an external white-noise process.

This is a spin model where the NV is coupled to a direction where the environment fluctuation and relaxation dynamics is suppressed, but that is mixed through Larmor precession with an axis to which the NV does not couple directly, but that experiences relaxation and fluctuations. We have shown thus that this spin model behaves exactly as the $m\neq0$ model we presented in the main text.

\section{Experimental Methods}\label{sec: methods}

\subsection{NV center experimental setup}
The NV centers used for our experiments are on nanopillar arrays \cite{Momenzadeh2014} on a diamond membrane. The membrane is mounted on a movable stage above a confocal microscope, which focuses a 520 nm green laser on the sample to excite the NV center. The position of the focal point can be controlled with nanometer resolution using galvo mirrors. The broad red-IR fluorescence (red of the zero phonon line at 637\,nm) is redirected to two single-photon counters using a dichroic mirror and a beam splitter. The photon counts are then read by a time tagger and an avalanche photodiode with a resolution of 300\,ps.

\subsection{Spin state control of the NV center}
The direction of the static magnetic field is controlled by manipulating the position of a permanent magnet above the diamond. To ensure a high fluorescence contrast, the direction of this field is chosen to align with the axis of symmetry of the NV center, determined by maximizing the difference in the frequencies of the $\left.|0\right\rangle \leftrightarrow \left.|1\right\rangle $ and $\left.|0\right\rangle \leftrightarrow \left.|-1\right\rangle $ transitions, which is evaluated using optically-detected magnetic resonance (ODMR) measurements.

The spin state of the NV center is manipulated by applying MW pulses created by mixing a MW signal with a signal from an arbitrary waveform generator (AWG). These pulses are then passed to a copper wire running above the surface of the diamond. All experiments are managed using Qudi\cite{Binder2017}, a Python-based software designed for carrying out quantum sensing measurements using NV centers.

\subsection{Characterization and control measurements on the NV} \label{sec:NV_characterization}

\begin{figure}[!ht]
    \centering
    \includegraphics[width=\columnwidth]{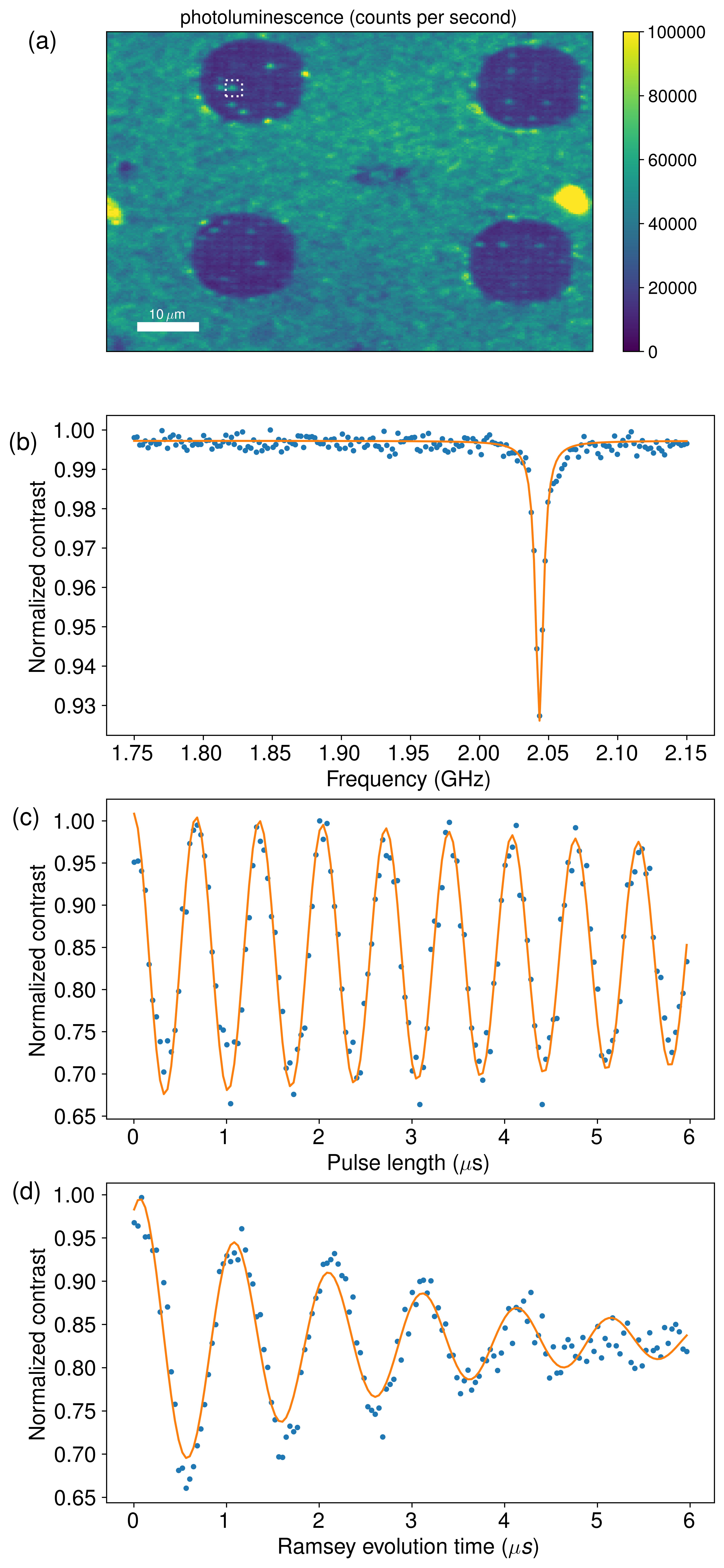} 
    \caption{(a) Confocal fluorescence microscope image showing the diamond used for the measurements presented in the main text. The dotted array are nanopillars fabricated to increase photon collection efficiency \cite{Momenzadeh2014}. Some of them are bright and host a single NV center. The one used here is marked by a white dotted square. (b) ODMR spectrum in the 1.75-2.15\,GHz range, showing the $\ket{0}\leftrightarrow \ket{-1}$ transition used for the measurements presented in panels (c) and (d). (c) Rabi measurement at the $\ket{0}\leftrightarrow\ket{-1}$ transition, showing a contrast of 34.1\% and a $\pi$ time of 344\,ns. (d) Ramsey measurement with a detuning of 1\,MHz, showing a $T_2^*$ of 3.52\,$\upmu$s.}
    \label{fig: ramsey_char}
\end{figure}

To see the effect of injected noise, one must first find an NV whose local noise environment has a sufficiently low amplitude, so that one can distinguish the effect of injected noise from the effect of environmental noise. After examining several NV defects, we found a suitable NV, which is marked with a white dotted rectangle in the fluorescence microscope image in Fig.\,\ref{fig: ramsey_char}(a). To characterize this NV, we performed ODMR and Rabi measurements, shown in Figs.\,\ref{fig: ramsey_char}(b), (c). These measurements were carried out after aligning the magnetic field with the NV axis by finding the magnet position which maximizes the ODMR splitting.  Based on the shift of the $\ket{0}\leftrightarrow\ket{-1}$ peak from the ZFS of 2.87\,GHz, we calculate the applied static magnetic field to be 298\,G. The Rabi measurement shows this NV to have a Rabi contrast of 34.1\% and a $\pi$ time of 344\,ns.  

\begin{figure}[!ht]
    \centering
    \includegraphics[width=\columnwidth]{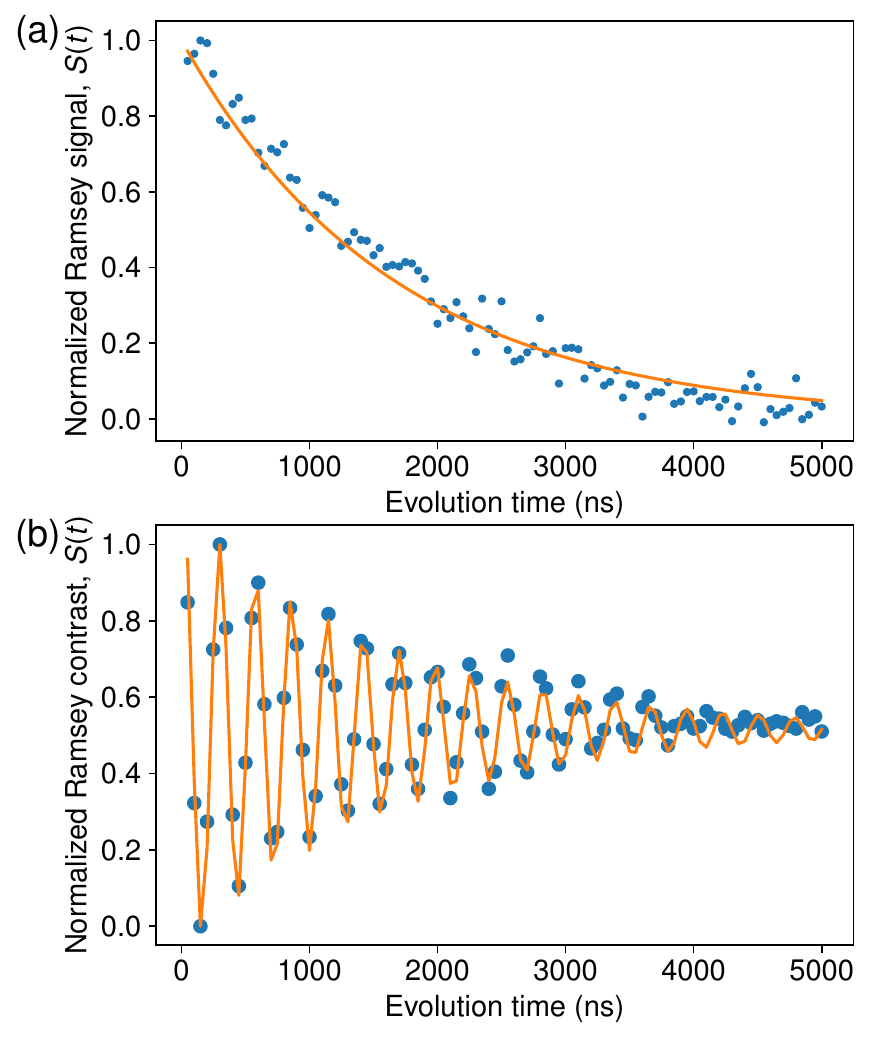}
    \caption{Control measurements for the NV: (a) Drift-corrected Ramsey signal, showing the $T_2^*$ without a detuning. (b) Ramsey signal with an injection of constant 0.45\,V signal, giving rise to a 3.6 MHz oscillation.}
    \label{fig: ramsey_control}
\end{figure}

\noindent Following this, we carry out two control experiments. To determine the effective $T^*_2$ of the NV, we carry out a (drift-corrected) Ramsey measurement with no injected noise, shown in Fig.\,\ref{fig: ramsey_control}(a). Assuming white noise from the environment, the $T^*_2$ fitted from this measurement is 1.65\,$\upmu$s. To estimate the coupling between the NV and the spiral, we also carry out a Ramsey measurement, where a constant potential of 0.45\,V is applied to the spiral (the maximum potential we can apply is 0.5\,V).  The Ramsey signal from this experiment, shown in Fig.\,\ref{fig: ramsey_control}(b), shows an oscillation at 3.6\,MHz, which is significantly faster than the decay rate of 0.608\,$\upmu$s${}^{-1}$, indicating that this NV is suitable for our injected noise measurements.

\subsection{Injection of noise and Ramsey decay measurements}
\label{sec:Injection_noise_Ramsey}
A realization of the desired type of noise is sent as an updated waveform to a Quantum Machines (QM) OPX, and the analog channel outputs the noise during the evolution time of the Ramsey sequence. A description of the noise generation algorithms for each noise type is given in Appendix \ref{sec:algorithms_noise_generation}. 

The noise is passed to a spiral antenna \cite{Yudilevich2023} below the diamond, with a hole in the center of the spiral to allow the green laser to excite any NV center in the exposed region of the diamond. Since we are using a spiral on the surface of the diamond to inject our noise, the magnetic field created by this current, measured in the center of the spiral, will always be perpendicular to the surface of the diamond, at some fixed angle with the z-axis of the NV. Therefore, all of the NV centers in this region experience a magnetic field proportional to the amplitude of the voltage supplied from the analog channel of the QM.

 To measure the effect of each type of injected noise on the Ramsey trace of our NV center, it is necessary, for each delay time, both to perform enough repetitions to appropriately measure the expectation value $\left\langle S_z\right\rangle $ after the Ramsey sequence and to inject enough realizations of the noise so that the $\left\langle S_z\right\rangle $ at each delay time will be sensitive to the full distribution of noise realizations. 
 
 To implement this, we carry out the measurement of each Ramsey trace using four nested loops.  The innermost loop repeats the Ramsey measurement 4$\cdot$10${}^{5}$ times for each realization at a given evolution time, the second loop repeats the first loop for 10 different realizations of noise for each evolution time, the third loop repeats the second loop for 100 different values of the evolution time, from 25\,ns to 2500\,ns, and the fourth loop repeats the third loop 10 times, for a total of 100 realizations per time point. The fourth loop is necessary so that any unexpected changes to the system will affect all time points. 
 
 Additionally, in order to minimize the effect of drift in the static magnetic field, every 30 minutes, the measurement cycle is paused, and a regular Ramsey measurement is performed with no injected noise, but with a 1\,MHz offset in the MW frequency. The Python program uses the results of this measurement to adjust the MW frequency to correct for this drift, and then measures another regular Ramsey measurement with a 1\,MHz offset to ensure that the correction was successful before continuing with the injected noise Ramsey measurements.

We note that, in principle, factors such as cross-talk, impedance mismatches, or antenna geometry could distort the injected noise. In our implementation, however, no systematic deviations attributable to these effects were observed: the measured Ramsey signals were accurately described by the analytical predictions across all regimes, including underdamped non-Markovian noise, where such distortions would be most apparent. If required, these effects could be further mitigated by characterizing the antenna transfer function and compensating for the injected noise spectrum accordingly. 

\subsection{Algorithms for generation of each noise type}
\label{sec:algorithms_noise_generation}
To implement the experiments, we generated discrete-time realizations of stochastic processes designed to reproduce the analytical correlation functions derived in Appendix~\ref{sec:Derivation_correlation_functions}. This section describes the algorithms used to produce each type of injected noise.

\subsubsection{Generation of Markovian noise trajectories} 

\begin{figure}[!ht]
    \centering
    \includegraphics[width=\columnwidth]{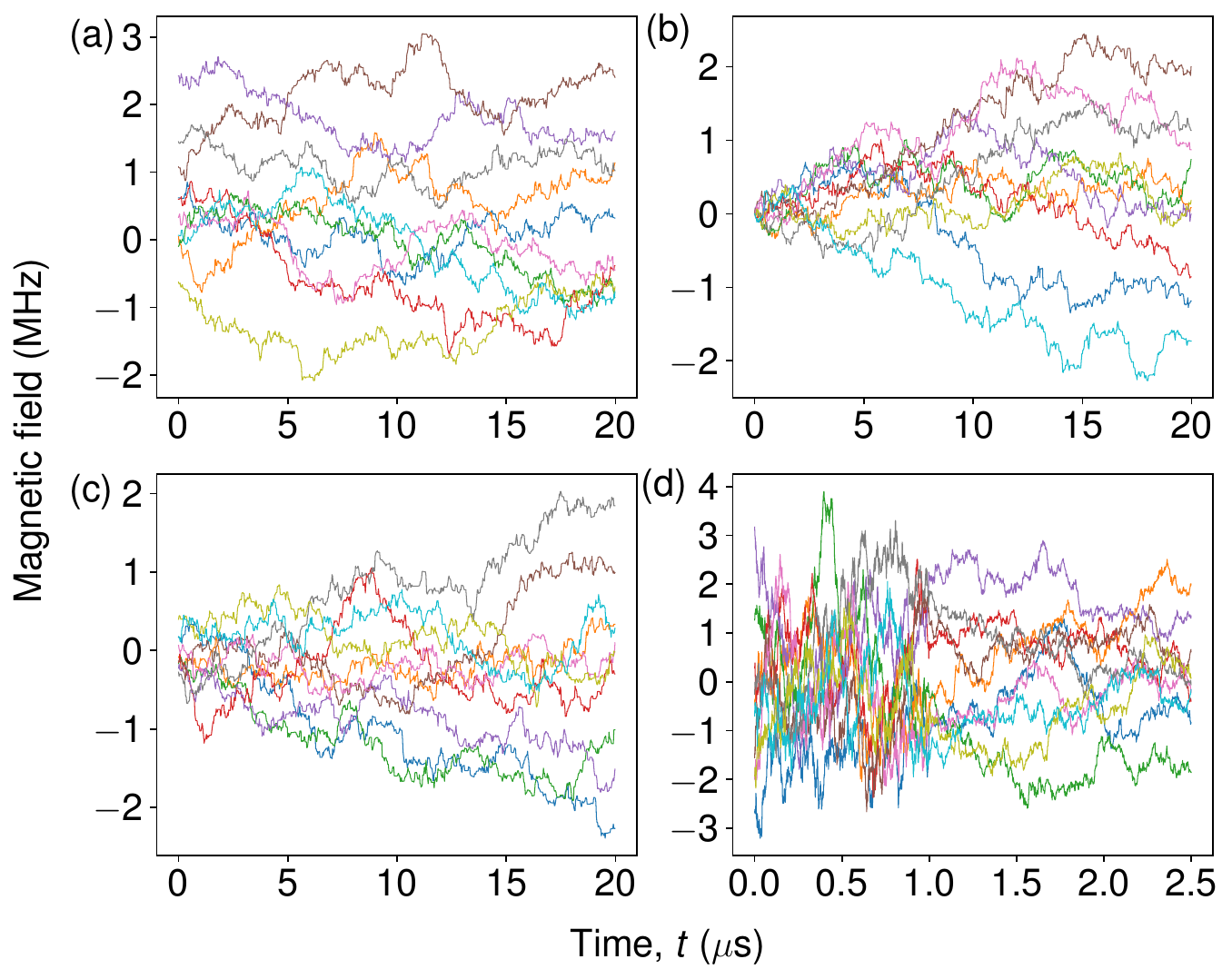}
    \caption{Simulation of the magnetization field as a function of time of  10 trajectories of Markovian noise with standard deviation 1.2\,MHz and correlation time, $t_c=42\,\upmu\text{s}$ (a) at equilibrium, (b) quenched, and (c) quenched with a delay of $t_d=2.5\,\upmu\text{s}$, and (d) 10 trajectories of Markovian noise with standard deviation 1.2\,MHz and a correlation time that switches abruptly from $t_c=200\,\text{ns}$ to $t_c = 2\,\upmu\text{s}$ at $t_s = 1\,\upmu\text{s}$.}
    \label{fig: trajectoriesM}
\end{figure}

 Because the correlation function for equilibrium Markovian noise has only one term, it is simple to generate this type of noise directly. As shown in Appendix \ref{sec:Markovian_correlation_functions}, Markovian noise has a correlation time given by $t_c$, and at equilibrium, the variance at each time point is constant $\left\langle {n(t)}^2\right\rangle ={\mathrm{\Delta }}^2$. In order to produce Markovian noise at a series of time points $0,\ \Delta t,\ 2\Delta t,\dots ,\ t_\text{total}-\Delta t$, we sample the first point from $N(0,{\mathrm{\Delta }}_M)$---the normal distribution with mean 0 and standard deviation ${\mathrm{\Delta }}_M$. For each subsequent point, we either retain the value of the previous point with probability $p_\text{ret}=e^{\frac{-\Delta t}{t_c}}$ or sample a new value from $N(0,{\mathrm{\Delta }}_M)$ with probability $1-p_\text{ret}=1-e^{\frac{-\Delta t}{t_c}}$. This produces a sequence of $\frac{t_\text{total}}{\Delta t}$ values, but this sequence will not represent continuous noise---rather, it will have large jumps roughly every $\frac{t_c}{\Delta t}$ points.  To produce continuous equilibrium Markovian noise with the desired parameters, we average $M=\frac{t_c}{\Delta t}$ of these sequences, setting ${\mathrm{\Delta }}_M=\mathrm{\Delta }\sqrt{M}$, so that the noise will change by a small amount (roughly $\mathrm{\Delta }\sqrt{\frac{2}{M}}$) at every point, but the standard deviation at each point will be $\mathrm{\Delta }$, and the correlation time $t_c$, as desired. An example of equilibrium Markovian noise generated using this method is shown in Fig.\,\ref{fig: trajectoriesM}(a). In our experiment, the temporal resolution is set by the pulser clock, with a discretization step of  $\Delta t = 4\,\mathrm{ns}$. The noise realizations are implemented electronically using a Quantum Machines OPX, which orchestrates and synchronizes the full experimental sequence---triggering the laser and microwave pulses and reading out the TTL signals from single-photon counting modules (avalanche photodiodes). The experimental setup has been described in detail in Ref. \cite{Zohar2023}.

For the quenched experiments, the quench is implemented digitally at the waveform-generation stage. 
Each stochastic sequence is produced with the required initial conditions—$n(0)=0$ for Markovian and 
$n(0)=n'(0)=0$ for non-Markovian quenches—and uploaded to the OPX arbitrary waveform generator. 
The OPX synchronizes the start of the noise waveform with the first $\pi/2$ pulse of the Ramsey sequence, 
ensuring that the quench onset coincides with the beginning of the sensor’s free-evolution period. 
The effective timing resolution of the noise control is 1–4 ns, much shorter than any correlation 
times $t_c$ explored experimentally, thereby guaranteeing that the quench dynamics are well resolved 
relative to the bath timescales.

A modification of this procedure can be used to generate quenched Markovian noise. As shown in Appendix \ref{sec:Markovian_correlation_functions}, quenched Markovian noise, which has the initial condition $n\left(0\right)=0$, has an extra term in the correlation function corresponding to the dissipation of this quench as the system returns to equilibrium. The result of this term is that the variance of the noise will change for each time point, so the variance ${\sigma (t)}^2$ of the normal distribution from which we sample must also change. From the correlation function $\left\langle n\left(t_1\right)n(t_2)\right\rangle ={\mathrm{\Delta }}^2\left(e^{\frac{-\left|t_1-t_2\right|}{t_c}}-e^{\frac{-(t_1+t_2)}{t_c}}\right)$, we have the condition 

\begin{equation}
\begin{aligned}
\left\langle n\left(t\right)n(t+\Delta t)\right\rangle &={\mathrm{\Delta }}^2\left(e^{\frac{-\Delta t}{t_c}}-e^{\frac{-(2t+\Delta t)}{t_c}}\right)\\
&=e^{\frac{-\Delta t}{t_c}}\left\langle n\left(t\right)n(t)\right\rangle 
\\&=p_\text{ret}\left\langle n\left(t\right)n(t)\right\rangle\,,
\end{aligned}
\end{equation}
so $p_\text{ret}=e^{\frac{-\Delta t}{t_c}}$, as before, and we also have the condition 
\begin{equation}
\begin{aligned}
\left\langle n\left(t+\Delta t\right)n(t+\Delta t)\right\rangle =&{\mathrm{\Delta }}^2\left(1-e^{\frac{-2(t+\Delta t)}{t_c}}\right)
\\=&p_\text{ret}\left\langle n\left(t\right)n(t)\right\rangle \\
+&(1-p_\text{ret}){\sigma (t+\Delta t)}^2\,,
\end{aligned}
\end{equation}

which gives $\sigma \left(t+\Delta t\right)=\Delta \sqrt{1-e^{\frac{-(2t+\Delta t)}{t_c}}}$. Because $\frac{\left\langle n\left(t\right)n(t+\Delta t)\right\rangle }{\left\langle n\left(t\right)n(t)\right\rangle }$ is independent of $t$, these same conditions apply for all time points. Thus, to generate quenched Markovian noise, we follow the same procedure as for equilibrium Markovian noise, averaging $M=\frac{t_c}{\Delta t}$ sequences, but with the modification that the first point of each sequence is always initialized to be 0, and for subsequent points, we either retain the value of the previous point with probability $p_\text{ret}=e^{\frac{-\Delta t}{t_c}}$ or we sample a new value from $N(0,{\sigma }_M'(t))$, where ${\sigma }'_M\left(t\right)={\sigma }_{eq}\sqrt{M}\sqrt{1-e^{\frac{-(2t+\Delta t)}{t_c}}}$. An example of quenched Markovian noise generated using this method is shown in Fig.\,\ref{fig: trajectoriesM}(b).

In addition to equilibrium and quenched Markovian noise, we also generate Markovian noise with a delayed quench and Markovian noise with a switched correlation time. To generate Markovian noise with a quench delay of $t_d$, we generate quenched Markovian noise for a series of time points $0,\ \Delta t,\ 2\Delta t,\dots ,\ t_\text{total}+t_d-\Delta t$ and discard the first $\frac{t_d}{\Delta t}$ time points. To generate Markovian noise with a correlation time that switches from $t_a$ to  $t_b$ at some time $t_s$, we generate equilibrium Markovian noise with a correlation time $t_a$ for a series of time points $0,\ \Delta t,\ 2\Delta t,\dots ,\ t_s-\Delta t$, and then we change $p_\text{ret}$ from $e^{\frac{-\Delta t}{t_a}}$ to $e^{\frac{-\Delta t}{t_b}}$ and continue generating noise for the rest of the time points $t_s,\ t_s+\Delta t,\ t_s+2\Delta t,\dots ,\ t_\text{total}-\Delta t$. Examples of each of these types of noise are shown in Fig.\,\ref{fig: trajectoriesM}(c), (d).

\subsubsection{Generation of underdamped non-Markovian noise trajectories}

\begin{figure}[!ht]
    \centering
    \includegraphics[width=\columnwidth]{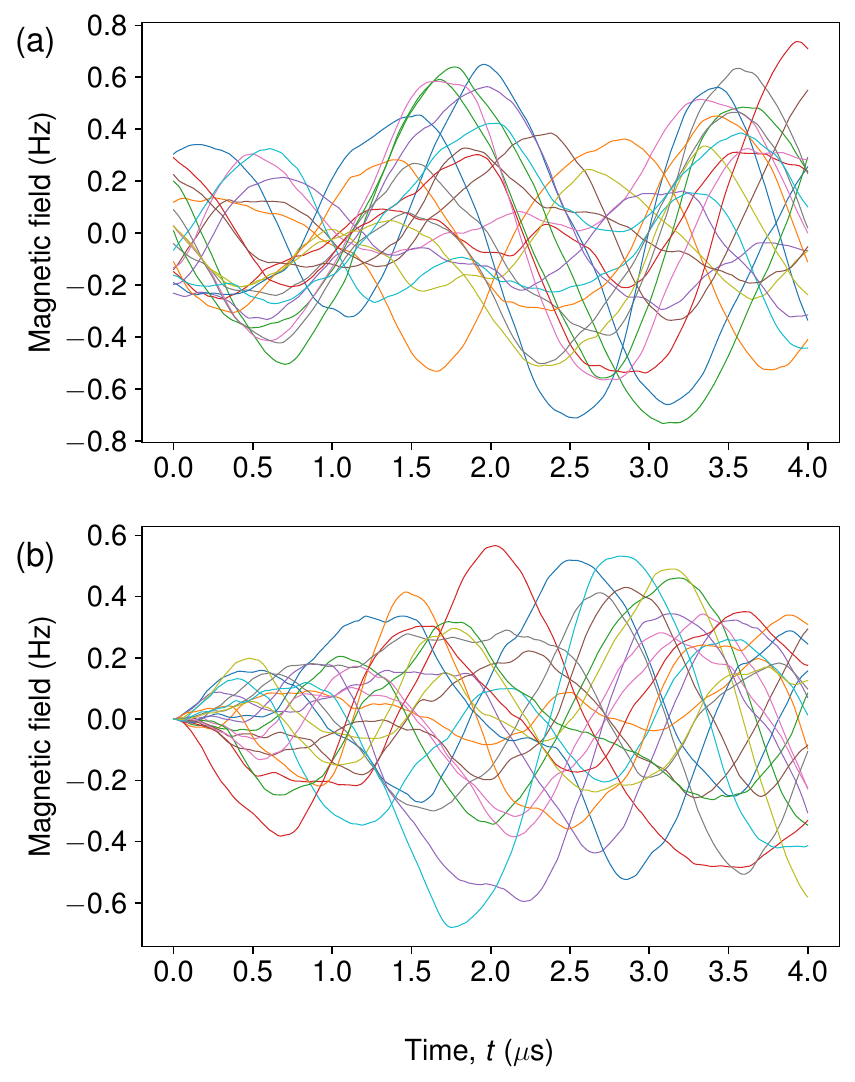}
    \caption{Simulation of the magnetization field as a function of time of 20 trajectories of underdamped non-Markovian ($m\neq 0$) noise with driving strength $A/m^2 = 0.45\cdot 10^{18}$ Hz$^5$, restoring frequency $\omega=3\,\text{MHz}$, and damping coefficient $\beta=0.5\,\text{MHz}$ (a) at equilibrium and (b) quenched.}
    \label{fig: trajectoriesNM}
\end{figure}

As discussed in Appendix \ref{sec:Underdamped_correlation_functions}, the correlation function for non-Markovian ($m\neq 0$) noise is more complex than for Markovian noise, so we cannot use the same strategy to generate non-Markovian noise directly.  Instead, we first generate some approximation of white noise $\eta(t)$, and then numerically solve for the response of an underdamped harmonic oscillator to this noise to obtain the trajectory $n(t)$.  We cannot reproduce the white noise correlation function $\left\langle \eta\left(t_1\right)\eta\left(t_2\right)\right\rangle =A\delta (t_1-t_2)$ exactly, so instead, we produce Lorentzian noise with a correlation function $\left\langle \eta\left(t_1\right)\eta\left(t_2\right)\right\rangle =\frac{A}{\Delta t}e^{\frac{-\left|t_1-t_2\right|}{\Delta t}}$ for some $\Delta t$ that will be sufficiently small so that the noise amplitude $n(t)$ we calculate at each point will not be affected much by the deviation from white noise.  We then numerically approximate
\begin{equation}    
n(t)-n_0(t)=\int^t_0{\frac{1}{m\Omega}\mathrm{sin}\mathrm{}(\Omega(t-s))e^{-\frac{t-s}{t_c}}\eta\left(s\right)ds}
\label{eq:Riemann_sum_nm_generation}
\end{equation}
for each time $t$ as a Riemann sum with spacing $\Delta t$.  For fully quenched non-Markovian noise, $n_0(t)=0$, and this is exactly the desired trajectory $n(t)$ of the noise. For stationary non-Markovian noise, we also independently sample $n'_0$ from the normal distribution $N(0,\mathrm{\omega_0 }\mathrm{\Delta })$ and $n_0$ from the normal distribution $N(0,\mathrm{\Delta })$, and then we compute $n_0(t)$ from Eq.\,(\ref{eq:G_hom_non_Mark}), and add it to the generated values of $n(t)-n_0(t)$ to give the trajectory $n(t)$ of the noise. Examples of equilibrium and quenched non-Markovian underdamped noise are shown in Fig.\,\ref{fig: trajectoriesNM}.

\subsection{Procedure for Ramsey measurements in the innermost loop} \label{sec:procedure_Ramsey_innermost_loop}

For all of the measurements presented here, the 520 nm laser pulses have a power of 2\,mW before the objective, and the Rabi driving power is 1.45\,MHz. The Ramsey measurements in the innermost loop begin with a 3 $\muup$s pulse from the 520 nm laser to initialize the NV center in the $\left.|0\right\rangle $ state. Following this, a $\frac{\pi }{2}$ pulse is applied at the resonant frequency of the $\left.|0\right\rangle \leftrightarrow \left.|-1\right\rangle $ transition of the NV, after which the evolution time $\tauup$ begins. During the evolution time, a noise trajectory of length $\tauup$ is injected using the spiral under the diamond, and the state of the NV center evolves to ${\left.|\psi \right\rangle }_{t=\tau }=\frac{1}{\sqrt{2}}\left(\left.|0\right\rangle +e^{i\varphi (\tau )}\left.|-1\right\rangle \right)$, where $\varphi \left(\tau \right)=\ {\gamma }_\mathrm{nv}B_0\tau +\gamma_\mathrm{nv}\int^{\tau }_0{\Delta B_z\left(t\right)dt}$. Afterwards, a $-\frac{\pi }{2}$ pulse is applied in order to convert the phase into a population difference between the states, and then a 1.6 $\upmu$s pulse from the 520 nm laser is applied. To maximize our ability to distinguish between the fluorescence of the $\left.|0\right\rangle $ and $\left.|-1\right\rangle $ states, we record the photon counts occurring between 220 $\upmu$s and 400 $\upmu$s after the start of the green laser pulse. We then repeat this sequence $4\cdot 10^{5}$ times for each realization of the noise, as specified in Appendix \ref{sec:Injection_noise_Ramsey}. Each Ramsey measurement in this innermost loop is repeated twice, once with a $-\frac{\pi}{2}$ pulse in the rotating frame before the readout and once with a $\frac{\pi}{2}$ pulse, and the Ramsey decay curve is given by the normalized difference between the counts from these two sets of experiments, so the reported signal from the Ramsey measurement is given by $S\left(t\right)=\frac{I_-\left(t\right)-I_+(t)}{{\mathrm{max} \left(I_-\left(t\right)-I_+(t)\right)\ }}$, where $I_-\left(t\right)$ denotes the total counts measured for the Ramsey measurements with evolution time $t$ and a $-\frac{\pi }{2}$ pulse before the readout, and $I_+\left(t\right)$ denotes the total counts measured for the Ramsey measurements with evolution time $t$ and a $\frac{\pi}{2}$ pulse before the readout.

\section{Validation of injected noise correlation functions} \label{sec:validation_injected_correlations}

To verify how accurately the injected noise reproduces the intended correlation functions, we generated 10000 realizations for each noise type---equilibrium Markovian and equilibrium underdamped non-Markovian noise---using typical experimental parameters.  For each case, we computed the empirical correlation function $\langle n(0)n(t)\rangle$ by averaging over realizations and compared it with the corresponding analytical prediction.

\begin{figure}[h!]
    \centering
    \includegraphics[width=0.4\textwidth]{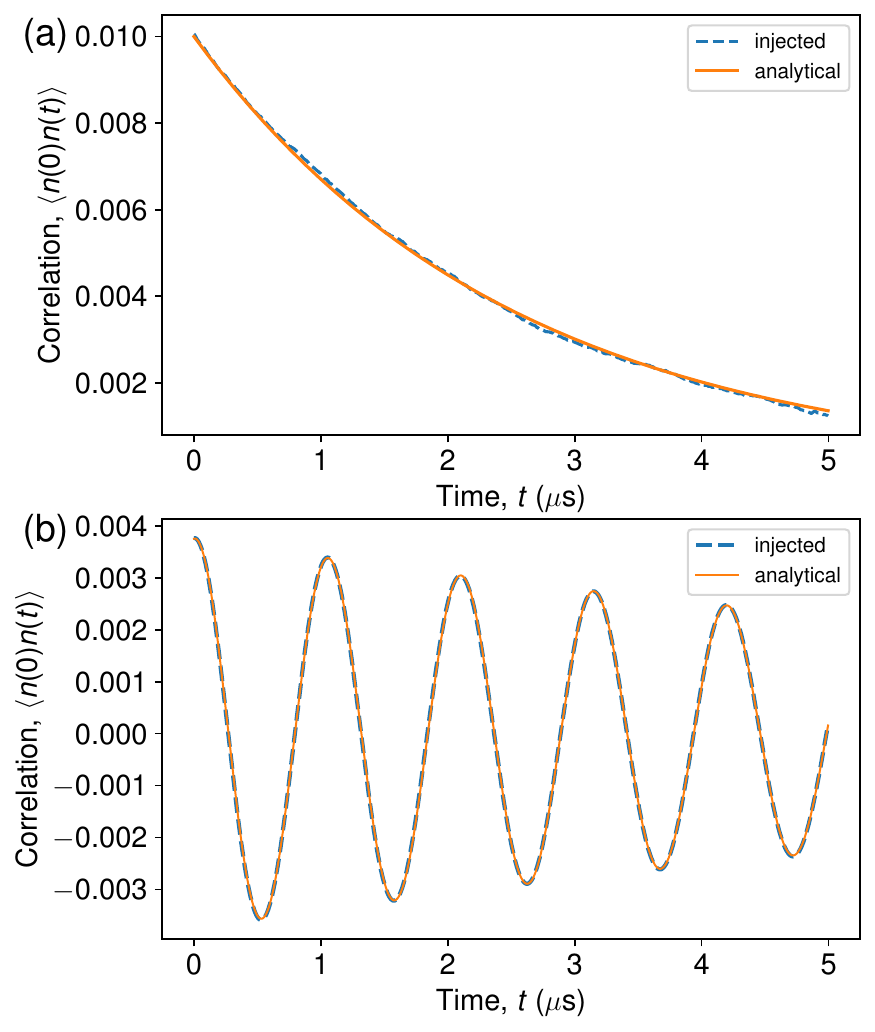}
    \caption{Comparison between analytically predicted (solid lines) and empirically computed (symbols) correlation functions obtained from 10000 realizations of (a) equilibrium Markovian noise with standard deviation $\Delta$ = 0.1\,V and correlation time $t_c = 2.5\,\upmu\text{s}$, and (b) equilibrium underdamped non-Markovian noise with driving strength $A/m^2 = 0.054\,\text{V}^2/\upmu\text{s}^3$, restoring frequency $\omega$ = 6\,MHz, and damping coefficient $\beta$ = 0.1\,MHz.}
    \label{fig:new10}
\end{figure}

As shown in Fig.\,\ref{fig:new10}, the empirical correlation functions $\langle n(0)n(t)\rangle$ closely match the analytically predictions across the entire relevant time range. For the generation of underdamped non-Markovian noise, the sampling rate must satisfy $\Delta t \ll 2\pi/\Omega$ to ensure convergence in the Riemann-sum representation in Eq. (\ref{eq:Riemann_sum_nm_generation}). 

In our experiments, we employ a discretization step of $\Delta t = 1$ ns for evaluating this integral, which is well below $2\pi/\Omega$ and therefore sufficient to guarantee convergence for all injected underdamped non-Markovian noise samples. At substantially higher frequencies (e.g., $\Omega$ in the GHz range), achieving convergence would require a finer discretization---i.e., a larger number of terms in the Riemann sum---leading to longer computation times.

\section{An additional example of switched Markovian noise} \label{sec:additional_example}

\begin{figure}[!htb]
    \centering
    \includegraphics[width=0.9\columnwidth]{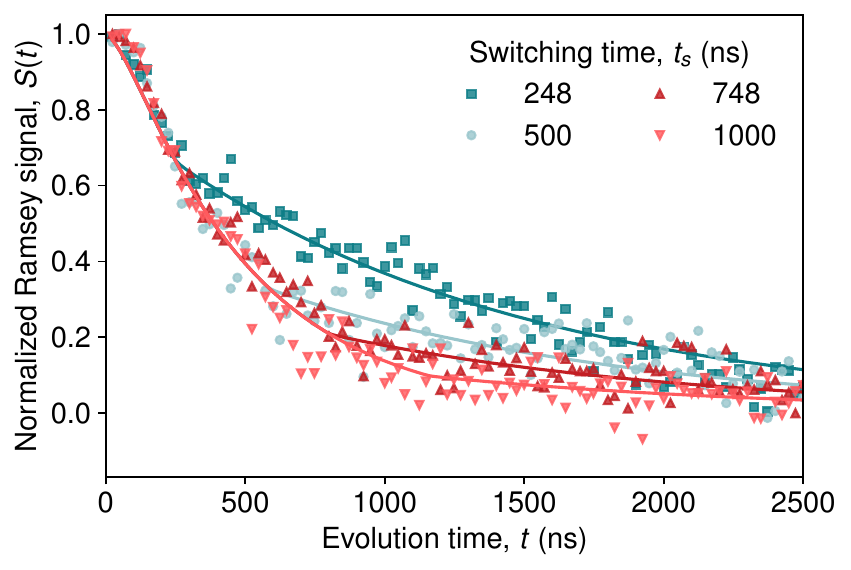}
    \caption{Ramsey signals measured for equilibrium Markovian noise with a standard deviation of 0.10\,V whose correlation time switches from $t_c=75$\,ns to $t_c=15$\,ns at times of $t_s=$248, 500, 748, and 1000\,ns.}
    \label{fig: switch_aux}
\end{figure}

We also measured Ramsey signals for equilibrium Markovian noise with a standard deviation of 0.10\,V and a correlation time that switches from $t_c=75$\,ns to $t_c=15$\,ns at times of $t_s=$248, 500, 748, 1000\,ns after the start of the measurement. The results are shown in Fig.\,\ref{fig: switch_aux}, along with fits using the analytical expression for the Ramsey signal of equilibrium Markovian noise with a switched correlation time derived in Appendix \ref{sec:underdamped_Ramsey_exponents}, with the time of the switch relative to the start of the measurement as a free parameter. The switching times extracted from the fits are shown in Table\,\ref{switch_table}, and match the known switching times used to generate the injected noise.

\begin{table}[!htb]
\begin{tabular}{|p{0.9in}|p{0.5in}|p{0.5in}|p{0.5in}|p{0.6in}|} \hline 
Injected switch time, $t_s$ (ns) & 248 & 500 & 748 & 1000 \\ \hline 
Switch time, $t_s$ (ns) & 239 $\mathrm{\pm}$ 18 & 576 $\mathrm{\pm}$ 29 & 790 $\mathrm{\pm}$ 44 & 1150 $\mathrm{\pm}$ 88 \\ \hline 
\end{tabular}
\caption{Comparison of the switch times extracted from fitting the Ramsey signals in Fig.\,\ref{fig: switch_aux} with the analytical expressions vs.\,the known values of the switch times for the injected noise.}
\label{switch_table}
\end{table}

\section{Parameter identifiability and complementary regimes}
\label{sec:Dependency_fitting_parameters}

The ability to extract all environmental parameters uniquely from experimental data depends on 
whether the measurement accesses the relevant dynamical regimes of the noise. 
In our framework, the parameters $t_c$, $A/m^2$, and $\omega_0$ (or $t_c$ and $A/\Gamma^2$ for Markovian noise) 
are formally independent. However, if the experiment probes only a restricted dynamical regime, the observable 
$\chi(t)$ may depend on specific parameter combinations, leading to apparent correlations in the fit. 
This is a universal feature of time-domain noise spectroscopy, not a peculiarity of our model.

For instance, in the stationary Markovian case, the short-time Ramsey attenuation depends only on the variance 
$\Delta^2 = \tfrac{1}{2} t_c A/\Gamma^2$, which combines $t_c$ and $A/\Gamma^2$. 
Hence, experiments restricted to this regime cannot determine these two parameters independently. 
Introducing a quench, however, modifies the short-time scaling to depend directly on $A/\Gamma^2$, 
thus providing an additional constraint that allows independent estimation of both $t_c$ and $A/\Gamma^2$. 
This demonstrates how complementary stationary and quenched experiments provide full parameter identifiability.

\begin{figure}
    \centering
    \includegraphics[width=\columnwidth]{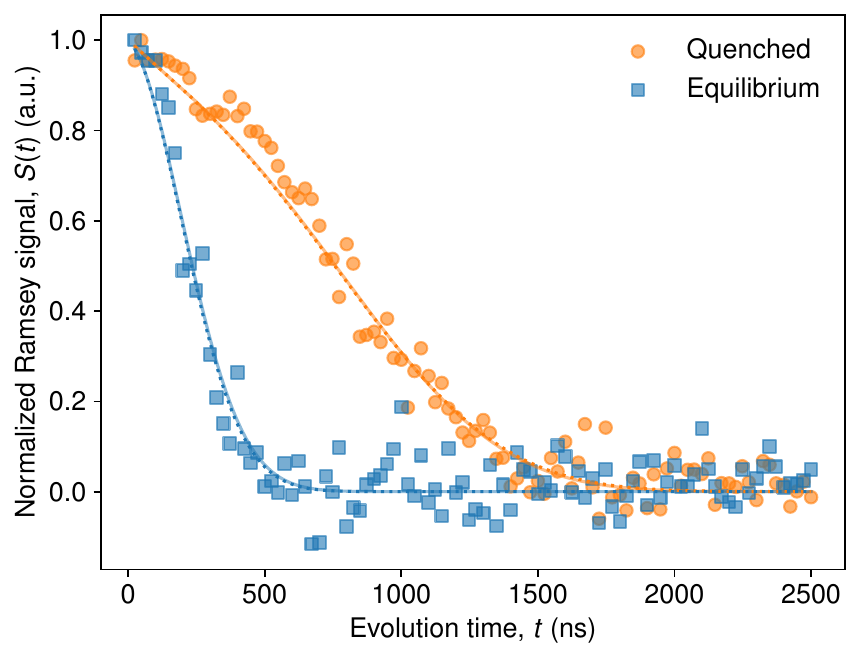}
    \caption{Independent and joint fits of equilibrium (blue) and quenched (orange) Markovian noise with standard deviation $\Delta = 0.10$ V and correlation time $t_c = 10\,\upmu\text{s}$. Transparent solid curves correspond to independent fits of each experiment. For the equilibrium case, the fit yields $t_c = 106 \pm 1552\,\upmu\text{s}$ with $t_\nu = 0.0686$ ($P(>t_\nu)=0.945$, dependency = 0.99999), which is not statistically significant, indicating that $t_c$ has no measurable influence on the curve within this regime ($t_\nu $ is Student's $t$-score). Dotted curves show a global fit to both datasets, giving $t_c = 10.53 \pm 0.85\,\upmu\text{s}$ with $t_\nu = 12.33$ ($P(>t_\nu)<0.00001$, dependency = 0.5048), a statistically significant result. The overlapping fitted curves demonstrate that both experiments are consistent with the same environmental parameters $\Delta$ and $t_c$. This comparison illustrates that a single experiment restricted to one dynamical regime cannot determine all noise parameters independently, whereas combining equilibrium and quenched measurements provides complete and unambiguous parameter estimation.}
    \label{fig:12}
\end{figure}

Figure~\ref{fig:12} illustrates this principle for equilibrium and quenched Markovian noise. 
Independent fits to each curve are over-parameterized, while a joint fit to both datasets 
recovers statistically meaningful and physically consistent values of $t_c$ and $\Delta$, 
confirming that combining different regimes resolves parameter correlations.

A similar reasoning extends to underdamped non-Markovian noise. 
In this regime, the environmental parameters influence 
the Ramsey response through distinct physical features: the short-time scaling of $\chi(t)$ encodes the noise strength $A/m^2$ (in a quenched experiment) and ${\mathrm{\Delta }}^2=\frac{At_c}{4{\omega_0 }^2m^2}$ (in a stationary one), 
the frequency of oscillatory revivals reflect $\omega_0$ in the underdamped regime, 
and their amplitude depends on $t_c$.
 Thus, by combining stationary and quenched experiments in the 
underdamped regime, one can disentangle $\frac{A}{m^2}$, $\omega_0$ and $t_c$— parameters that may otherwise appear 
correlated when only a single regime is probed. 
This demonstrates the broader applicability of our framework for identifying and isolating 
environmental parameters across both Markovian and non-Markovian dynamics.

\bibliography{main}

\end{document}